\documentclass{pasa}%

\usepackage{graphicx}

\title[Dynamic formation of black hole binary mergers]{Dynamically Formed Black Hole Binaries: In-Cluster Versus Ejected Mergers}

\author[Anagnostou et al.]{O. Anagnostou$^1$\thanks{Contact: oanagnostou@student.unimelb.edu.au}, M. Trenti$^{1,2}$ and A. Melatos$^{1,2}$
\affil{$^1$School of Physics, The University of Melbourne, VIC 3010, Australia}%
\affil{$^2$Australian Research Council Centre of Excellence for Gravitational Wave Discovery (OzGrav),\\
University of Melbourne, Parkville, VIC 3010, Australia}
}%

\jid{PASA}
\doi{10.1017/pas.\the\year.xxx}
\jyear{\the\year}

\usepackage{aas_macros}
\usepackage{hyperref} 
\usepackage{amsmath}
\usepackage{amssymb}

\usepackage{verbatim} 
\usepackage{xcolor}
\usepackage{subfig}
\usepackage{lipsum}
\usepackage{threeparttable}
\usepackage{natbib}
\usepackage{changepage} 

\hypersetup{colorlinks,citecolor=blue,linkcolor=blue,urlcolor=blue}

\hypersetup{draft}

\graphicspath{{Figures/}}

\begin{document}

\begin{frontmatter}
\maketitle

\begin{abstract}
The growing number of black hole binary (BHB) mergers detected by the Laser Interferometer Gravitational-Wave Observatory (LIGO) have the potential to enable an unprecedented characterization of the physical processes and astrophysical conditions that govern the formation of compact binaries. In this paper, we focus on investigating the dynamical formation of BHBs in dense star clusters through a state-of-art set of 58 direct N-body simulations with N $\leqslant200,000$ particles which include stellar evolution, gravitational braking, orbital decay through gravitational radiation and galactic tidal interactions. The simulations encompass a range of initial conditions representing typical young globular clusters, including the presence of primordial binaries. The systems are simulated for $\sim 12$ Gyr. The dataset yields 117 BHB gravitational wave events, with 97 binaries merging within their host cluster, and 20 merging after having been ejected. Only 8\% of all ejected BHBs merge within the age of the Universe. Systems in this merging subset tend to have smaller separations and larger eccentricities, as this combination of parameters results in greater emission of gravitational radiation. We confirm known trends from Monte Carlo simulations, such as the anti-correlation between the mass of the binary and age of the cluster. In addition, we highlight for the first time a difference at low values of the mass ratio distribution between in-cluster and ejected mergers. However, the results depend on assumptions on the strength of gravitational wave recoils, thus in-cluster mergers cannot be ruled out at a significant level of confidence. A more substantial catalogue of BHB mergers and a more extensive library of N-body simulations are needed to constrain the origin of the observed events.

\end{abstract}

\end{frontmatter}

\section{INTRODUCTION}
\label{sec:intro}

 In 2016, the Laser Interferometer Gravitational-Wave Observatory (LIGO) Scientific Collaboration (LSC) announced the first detection of gravitational waves (GWs), produced by a Black Hole Binary (BHB) merger \citep{2016PhRvL.116f1102A}. The detection sparked a new era for astronomy. GW astronomy has been used to measure BH properties more accurately than ever before, including spins, masses and BHB merger rates. At the end of LIGO's second observing run (O2) there have been 11 reported GW events, 10 from BHB mergers and 1 from a binary neutron star merger \citep{2016PhRvL.116f1102A,PhysRevX.6.041015,PhysRevLett.116.241103,PhysRevLett.118.221101,Abbott_2017,2018arXiv181112907T,stoyan_binnewies_friedrich_2008}. With the completion of the O3 run on March 27 2020, the first few GW BHB detections have been published \citep{2020ApJ...896L..44A, 2020arXiv200408342T, 2020ApJ...892L...3A}, and there will soon be an expanded catalogue of BHB merger detections which will allow for detailed comparisons with theoretical models, reportedly including a total of 56 GW detections \citep{LIGOnews}.

The origin of merging BHBs is still mostly unknown. There are two broad formation channels: common envelope evolution, and dynamical evolution. In the common envelope evolution channel, the BHB progenitors form together in a stellar binary system and co-evolve \citep{2002ApJ...572..407B,2016Natur.534..512B,2012ApJ...759...52D,2016ApJ...819..108B,2013ApJ...779...72D}, isolated from gravitational interactions with other stellar objects. In the dynamical evolution channel, the BHB forms via gravitational interactions with other stellar bodies \citep{2006ApJ...637..937O,2010MNRAS.402..371B,2007PhRvD..76f1504O,2013MNRAS.435.1358T,2018MNRAS.480.2343C}, with evolution strongly dependent on dynamical processes within dense environments, like young star clusters or globular clusters (GCs) \citep{Baumgardt_2003}.

\paragraph{}
In order to better understand the formation of merging BHBs, work has been done to model the evolution of compact binaries based on the current understanding of stellar formation and evolution. Of particular interest is the modelling of dense star clusters through both direct N-body and Monte Carlo (MC) methods \citep{2015ebss.book..225M,2010MNRAS.402..371B,2017MNRAS.469.4665P, 2016PhRvD..93h4029R,2019arXiv190610260R, 2020IAUS..351..438G, 2020IAUS..351..395A, 2018MNRAS.480.5645H, 2018MNRAS.478.1844A, 2018arXiv180208654S, 2020ApJ...894..133A, 2018MNRAS.481.5445S, 2019arXiv190605864A, 2020ApJS..247...48K, 2019PhRvD.100d3027R, 2019PhRvD..99f3003K, 2020MNRAS.492.2936A}. GCs naturally provide the dense stellar environment and low metallicities required to produce high mass BH mergers through the dynamical evolution channel \citep{2018arXiv180605820M}. \citet{2010ApJ...715L.138B} even proposed that low-metallicity massive star progenitors dominate the BHB merger rate. Metallicity effects the progenitor mass and the  mass loss rate, as (1) cooling is suppressed in low metallicity gas clouds, leading to little fragmentation \citep{2005fost.book.....S,2018arXiv180711489C}, and (2) low metallicity stars also experience less extreme stellar winds because of their lower opacity, resulting in lower mass loss and heavier stellar remnants \citep{2003ApJ...591..288H}. Recent work has also focused on modelling BHB mergers in open clusters, which tend to be smaller and younger than GCs \citep{2017MNRAS.467..524B,2018MNRAS.473..909B,2018MNRAS.481.5123B,banerjee2020stellarmass, 2019MNRAS.483.1233R, 2019MNRAS.487.2947D, 2020MNRAS.tmp.2334D}. \citet{2017MNRAS.467..524B} used direct N-body simulations of young, massive open clusters to investigate BHB formation. They found a prevalence for in-cluster mergers mediated by triple-body interactions, as opposed to systems merging after ejection from their host cluster. 

\paragraph{}
 Separate but related work has focused on nuclear star clusters, dense, luminous star clusters within galactic nuclei. Modelling of dynamical formation of BHBs within nuclear star clusters provides insight into another potential source of GW events \citep{2016ApJ...831..187A, 2020arXiv200614632F, 2018ApJ...856..140H, 2020arXiv200601867F, 2020IAUS..351...80D}. The unique properties of nuclear star clusters allow for different pathways leading to GW events. For example, \citet{2020ApJ...897...46F} propose a formation channel for intermediate mass BH mergers with stellar BHs through gravitational wave capture.

 \paragraph{}
As GCs evolve, heavier stars segregate toward the core in an attempt to establish partial energy equipartition \citep{2013MNRAS.435.3272T}, and BHs experience frequent strong dynamical interactions among themselves and with other massive bodies. These include various three and four-body interactions such as hardening and exchange, leading to tighter binaries and more mergers. Dynamically formed BHBs differ from isolated binaries in that they can be driven to measurable eccentricities when entering the LIGO frequency band \citep{2018PhRvD..98l3005R}. The high central densities of GCs make it also possible to dynamically form hierarchical triplet systems, three-body systems with a tight inner binary and wide companion, with an expectation of one triple every $\sim 100$ binaries \citep{2008MNRAS.387..815T}. Through Kozai-Lidov oscillations \citep{1962P&SS....9..719L,1962AJ.....67..591K}, angular momentum can be transferred between inner and outer orbits, driving the inner orbit to significant eccentricities at the expense of the outer orbital inclination \citep{1962AJ.....67..591K}. As a result, a near-circular BHB which otherwise would not merge can be driven to high eccentricities, leading to efficient gravitational radiation, rapid orbital decay, and hence a merger \citep{2018arXiv180711489C}. If the gravitational radiation cannot effectively circularize the system, then the  Kozai-Lidov oscillations can push the binary above the 10 Hz LIGO cutoff with measurable eccentricity \citep{2012ApJ...757...27A,2013PhRvL.111f1106S}. It is also possible for a single-single or binary-single scattering event to form eccentric binaries that produce eccentric mergers through gravitational capture \citep{2018PhRvD..97j3014S,2019arXiv190711231S}.

\paragraph{}
In this paper, we resort to a new large set of direct N-body simulations designed to investigate properties of compact object mergers \citep{2019MNRAS.485.5752D, 2016ApJ...819...70M} to focus on the formation of BHBs within mid-sized star clusters containing $5\times 10^4 \leqslant N \leqslant 2\times10^5$ stars, a range that includes the $~ 8.1\times10^4 M_{\odot}$ median mass for Milky Way GCs (see \citealt{2003gmbp.book.....H} Table 1.1) under a typical stellar initial mass function even after accounting for mass loss during dynanical evolution. We compare the properties of binaries that merge within a cluster to systems that merge after they are ejected from the cluster, with the goal of understanding how the channels are represented in current and future LIGO observations. Our set of 58 direct N-body simulations explore a range of cluster structure and initial conditions. They are carried out using NBODY6 \citep{2003gnbs.book.....A} with GPU support \citep{2012MNRAS.424..545N}, an efficient code which has been used extensively in the modelling of GCs \citep{2010ApJ...708.1598T,2017MNRAS.469.4665P,2017MNRAS.467..524B}. NBODY6 includes subroutines for single and binary stellar evolution, and dynamical interactions are incorporated directly through the equations of motion (see Section \ref{subsec:nbody6}), offering distinct advantages for accuracy over approximate methods such as MC algorithms. The simulations also incorporate advanced prescriptions for relativistic effects.

\paragraph{}
In Section \ref{sec:method}, we describe the N-body code used in modelling the GCs, and the specific cluster models and initial conditions. In Section \ref{sec:in-cluster}, we explore the in-cluster BHB mergers and the distribution of their parameters. In Section \ref{sec:ejected}, we compare and contrast the properties of the in-cluster mergers to the BHBs that merge after being ejected from their host cluster. Finally, in Section \ref{sec:ligo}, we compare our results to the 10 BHB mergers detected during LIGO O1 and O2 and explore the implications for their formation and for future detections. 
\section{METHOD}
\label{sec:method}

\subsection{NBODY6} 
\label{subsec:nbody6}
N-body simulations directly integrate the equations of motion \citep{2003gmbp.book.....H}. Unlike MC methods, they include all gravitational interactions to the desired numerical accuracy \citep{2015ebss.book..225M,2003gnbs.book.....A}, without relying on simplifications of the dynamics (e.g. softening) or specific assumptions for the cross sections of close encounters. Achieving such computational accuracy comes with increased computational cost, so that the particle number we can effectively follow with current hardware in this study is limited to N $= 50$K, $100$K and $200$K stars (where K$\equiv 1000$). The \texttt{NBODY} series of codes, with NBODY6 \citep{1999PASP..111.1333A} used in our study, have been specifically designed for the modelling of star clusters. NBODY6 employs block time step schemes and algorithmic regularization of multi-star systems \citep{2003gnbs.book.....A,springerbook,2006NewA...12..124M}, resulting in significant improvements to the computational efficiency compared to similar direct N-body integrators. 

NBODY6 implements single and binary stellar evolution through single and binary stellar evolution algorithms ("SSE" and "BSE" respectively, see \citealt{2000MNRAS.315..543H,2013ascl.soft03014H}), which have been designed to handle complex evolutionary processes, including mass transfer, common envelope evolution, collisions and supernova kicks. General Relativistic (GR) effects are also implemented through the Peters orbital evolution equations \citep{1963PhRv..131..435P} and Post Newtonian (PN) terms through BSE. This allows for accurate treatment of close encounters of compact bodies, including gravitational radiation. The version of NBODY6 we use was forked from the official branch in 2015 and customized for increased accuracy in the sphere of influence of a BH (see \citealt{2010ApJ...708.1598T} for further details). Note that our version does not include the recently implemented upgrades to some stellar evolution prescriptions for massive stars (stellar wind, mass fallback and pair-instability supernova) presented in \citet{2019arXiv190207718B}. While we do not expect significant changes to our key results as these are relatively minor changes to an already complex code, we plan to run a new set of simulations as a follow up to this work using the code resulting from the merger of our customisations and the current NBODY version.

\paragraph{}
For compact binary systems located inside the simulated star cluster, merger times are calculated by the BSE algorithm directly. For BHBs that are ejected from the cluster before merging, inspiral times are calculated in post-processing using the orbit-averaged Peters equation, since NBODY6 stops computing stellar evolution for particles ejected from the system. The inspiral time in these cases can be approximated by the integral \citep{PhysRev.136.B1224}

\begin{eqnarray} \label{eq1}
    & & T_{\text{insp}}(a_0,e_0, M_1, M_2) = \frac{12}{19}\frac{{c_0}^4}{\beta} \nonumber \\ 
    & & \times \int_0^{e_0}\frac{e^{29/19}\left(1 + \frac{121}{304}e^2\right)^{1181/2299}}{(1 - e^2)^{3/2}}de,
\end{eqnarray}

\noindent with $\beta = \frac{64}{5}\frac{G^3}{c^5}M_1M_2\left(M_1 + M_2\right)$, where $M_1$ and $M_2$ are the primary and secondary masses respectively, $a_0$ and $e_0$ are the semi-major axis and eccentricity respectively upon ejection from the cluster, and $c_0$ is a constant fixed by $a_0$ and $e_0$, given by \citep{PhysRev.136.B1224}

\begin{equation} \label{eq:c}
    c_0(a_0,e_0) = a_0\left[\frac{(1-e_0^2)}{e_0^{12/19}}\left(1 + \frac{121}{304}e_0^2\right)^{-870/2299} \right].
\end{equation}

\subsection{Gravitational Recoil}
\label{subsec:Gravitational_Recoil}

Depending on the orientation of a compact binary and the spins of its components, merger remnants can experience significant recoil due to the asymmetric emission of GWs \citep{PhysRev.128.2471,1973ApJ...183..657B,1984MNRAS.211..933F}. Full calculations for gravitational recoil require numerical relativity. Our simulations do not include recoil kicks to remnants from black hole mergers computed on the fly, as this was not implemented in NBODY6 before initial submission of this paper \citep{2006ApJ...637..937O,2010MNRAS.402..371B,2014MNRAS.440.2714B,2013MNRAS.435.1358T,2017MNRAS.469.4665P}\footnote{We note that there has been recent work in implemented gravitational recoil into direct NBODY codes through numerical relativity \citep{2020arXiv200407382B}, while this paper was under peer-review.}. A caveat is that numerical modelling has shown that the recoil speed may exceed $1000~\text{kms}^{-1}$ in special cases \citep{2007ApJ...659L...5C}, and so some of our results may be impacted by the lack of gravitational recoil physics. Future simulations will explore the feasibility of including this additional physics to improve the realism of our numerical modeling. 

\paragraph{}
In order to at least partially account for gravitational recoils in our simulations, we apply an approximate treatment of this additional physical ingredient in post processing. For this, we use an analytic approximation for the recoil velocity, dependent on the symmetric mass ratio presented in \citet{2006PhRvD..74l4010S}:

\begin{equation}
    a\eta^2\sqrt{1 - 4\eta^2}\left( + b\eta+ c\eta^2\right),
\end{equation}\

where $\eta = \frac{M_1M_2}{(M_1 + M_2)^2}$ is the symmetric mass ratio, and $a,b$ and $c$ are free parameters. \citet{2006PhRvD..74l4010S} presented multiple models to fit the free parameters. We implement a fit in between the lower and upper bound, corresponding to $a = 9082$, $b = -1.43$ and $c = 1.68$. Using this analytic approximation, we calculate the gravitational velocity kick for all in-cluster BHB mergers, and determine whether the merger remnant is ejected from its cluster as a result of the kick. Using this information, we construct a refined set of BHB mergers trees which exclude any binaries containing a remnant BH flagged to have been ejected due a previous merger kick. We refer to this set of BHBs as retained In-Cluster Mergers. Note as a caveat that this analytic model is merely an approximation, and among other aspects does not account for BH spins. Also, the BHs flagged as ejected are retained in the system and more likely to interact with other BHs because of their higher mass, hence this approach might be too conservative in removing merger events from our analysis, but it nevertheless provides an indication of the likely impact of the recoils on our results. For completeness, we also consider in our analysis the baseline scenario where recoils are ignored.

\subsection{Cluster Models}
\label{subsec:models}
The simulations in this paper are a combination of 33 runs from \citet{2019MNRAS.485.5752D}, which have been augmented by an additional 25 simulations for a total of 58 realizations, which are summarized in table \ref{tab:models}. In the current paper, we analyze the population of BHB mergers produced by these simulations. We organise the simulations into three main groups based on initial particle number with  $N = 50$K (11 simulations), $100$K (42 simulations) and $200$K (5 simulations). The majority of the simulations are initialized with $100$K particles, as this represents the best compromise between run time and realistic cluster size. Metallicity, primordial binary fraction, stellar remnant natal kick distribution, half mass radius, initial mass function, dimensionless potential and the presence of an Intermediate Mass BH (IMBH) vary across simulations to explore a wide range of parameter space. For some models, multiple independent realizations of the same set of parameters are followed to probe run-to-run variations.

Three metallicities are considered: $0.083Z_{\odot}$, $0.16Z_{\odot}$ and $1.6Z_{\odot}$. The  $0.083Z_{\odot}$ and $0.16Z_{\odot}$ models are in the middle to upper range of the Milky Way GC metallicity distribution \citep{1996AJ....112.1487H}. The $1.6Z_{\odot}$ model is more metallic than Milky Way clusters and is only used in one simulation to test the effects of high metallicity on cluster evolution. Although the LIGO GW events do not come from Milky Way clusters, the distribution of extra-galactic GC metallicities is not well constrained, and so we cannot make comparisons to our model metallicities. Exploration of the impact of further reductions in the initial metallicity will be devoted to future work. 

\paragraph{}
We consider clusters with initial half mass radii ($r_h$) of $1.5,2.5,4$ and $6$pc. All models are assumed to be spherically symmetric, with initial conditions drawn from a \cite{1966AJ.....71...64K} distribution. The spherical density profile, $\rho(r)$, is characterized by a dimensionless concentration parameter: 

\begin{equation} \label{eq2}
    W_o = \frac{|\phi_o|}{\sigma_o^2},
\end{equation}

\noindent where $\sigma_o$ is the central velocity dispersion, and $\phi_o$ is the corresponding central potential. We consider systems with $W_o = 3,5$ and $7$, representing relatively extended systems to relatively concentrated systems \citep{2003gmbp.book.....H}. Most simulations have $W_o = 7$ as this is a value close to the long-term concentration of typical simulated star clusters, which is reached after the system relaxes irrespective of the initial concentration (see \citealt{2007MNRAS.374..344T}).

\paragraph{}
Particle masses are drawn from either a Kroupa \citep{2001MNRAS.322..231K} or Salpeter \citep{1955ApJ...121..161S} initial mass function (IMF) of the form: 

\begin{equation} \label{eq3}
    P(m)dm \sim m^{\alpha}dm, 
\end{equation}

\noindent with

\begin{equation} \label{eq4}
    \alpha = - 2.35
\end{equation}

\noindent for the Salpeter IMF, and

\[ \alpha = \begin{cases}  
        
      -0.3 & M < 0.08M_{\odot} \\
      -1.3 & 0.08M_{\odot}\leq M \leq 0.5M_{\odot} \\
      -2.3 & 0.5M_{\odot} \leq M
   \end{cases}
\]

\noindent for the Kroupa IMF. Regardless of IMF, we generate particle masses in the range $ 0.08M_{\odot}$ to $100M_{\odot}$. In addition to regular particles drawn from an IMF, a small subset of models contain an IMBH. The IMBHs are generated with a mass of either $100M_{\odot}$, $200M_{\odot}$ or $400M_{\odot}$, corresponding to $0.15-0.3\%$ of the initial cluster mass, and hence either begin as or rapidly become the most massive particle in the cluster. We initialize the IMBHs as static (zero kinetic energy) particles in the centre of the potential well, and allow them to interact and merge with other particles so that they wander within the host cluster; see \cite{2016ApJ...819...70M} and  \cite{2018MNRAS.475.1574D} for more details. Although mergers between IMBHs and Stellar-mass black holes (SMBHs) do occur in the simulations, we do not include them in our analysis and refer instead to \citet{2016ApJ...819...70M}.

\begin{table*}
\centering
\caption{Summary of N-body simulations. For each simulation (identified by a unique ID) we report (from left to right) the initial number of stars; the initial IMBH mass in $M_\odot$; the velocity dispersion of the natal kick imparted to stellar remnants $\sigma_k$, normalized to the initial cluster velocity dispersion $\sigma_*$; the fraction of primordial binaries $f$; the metallicity $Z$; the number of distinct realizations of the same initial conditions ($N_\mathrm{sim}$); the initial half-mass radius $r_\mathrm{h,0}$ in pc; the initial dimensionless potential of the King Model $W_0$ and the initial mass function (taken from either \citealt{2001MNRAS.322..231K} or \citealt{1955ApJ...121..161S}).}
\label{tab:models}
\begin{threeparttable}[b]
\begin{tabular}{|l|c|c|c|c|c|c|c|c|c|}
\hline
\text{ID} & \text{N} &$M_{\mathrm{bh},0}$ &$\sigma_k/\sigma_*$ &$f$ &$Z$ &$N_\mathrm{sim}$  &$r_\mathrm{h,0}$ &$W_0$  &IMF\\
\hline
\text{can50k}   & 50k	&-		& 1.0	&-		&0.002		&4  &2.5 &7 &Kr\\
\text{fb1050k}  & 50k	&-		& 1.0	&0.10	&0.002		&1  &2.5 &7 &Kr\\
\text{IMBH50k}  & 50k	&100	& 1.0	&-		&0.002		&1  &2.5 &7 &Kr\\
\text{Z50k}	    & 50k	&-		& 1.0	&-		&0.001		&2  &2.5 &7 &Kr\\
\text{kick50k}  &50k	&-		& 2.0	&-		&0.002		&1  &2.5 &7 &Kr\\
\text{rh450k}   &50k	&-		& 1.0	&-		&0.002		&1  &4 &7 &Kr\\
\text{rh50k}    &50k	&-		& 1.0	&-		&0.002		&1  &6 &7 &Kr\\
\hline
\text{can100k}  & 100k	&-		& 1.0	&-		&0.002		&7  &2.5 &7 &Kr\\
\text{no\_sse\footnotemark[1]} & 100k	&-		& 1.0	&-		&0.002		&1  &2.5 &7 &Kr\\
\text{fb01}     & 100k	&-		& 1.0	&0.01	&0.002		&1  &2.5 &7 &Kr\\
\text{fb03}	    & 100k	&-		& 1.0	&0.03	&0.002		&1  &2.5 &7 &Kr\\
\text{fb05}     & 100k	&-		& 1.0	&0.05	&0.002		&2  &2.5 &7 &Kr\\
\text{fb07}     & 100k	&-		& 1.0	&0.07	&0.002		&1  &2.5 &7 &Kr\\
\text{fb10}	    & 100k	&-		& 1.0	&0.10	&0.002		&2  &2.5 &7 &Kr\\
\text{imbh}	    & 100k	&100	& 1.0	&-		&0.002		&1  &2.5 &7 &Kr\\
\text{IMBH}	    & 100k	&200	& 1.0	&-		&0.002		&1  &2.5 &7 &Kr\\
\text{Z}	    & 100k	&-		& 1.0	&-		&0.001		&1  &2.5 &7 &Kr\\
\text{highZ}	& 100k	&-		& 1.0	&-		&0.02		&1  &2.5 &7 &Kr\\
\text{kick}	    & 100k	&-		& 2.0	&-		&0.002		&1  &2.5 &7 &Kr\\
\text{rh}    & 100k	&-		& 1.0	&-		&0.002		&1  &6 &7 &Kr\\
\text{rh09}    & 100k	&-		& 1.0	&-		&0.002		&1  &0.9 &7 &Kr\\
\text{rh1}   & 100k	&-		& 1.0	&-		&0.002		&1  &1.5 &7 &Kr\\
\text{rh4}   & 100k	&-		& 1.0	&-		&0.002		&1  &4 &7 &Kr\\
\text{IMF}   & 100k	&-		& 1.0	&-		&0.002		&1  &2.5 &7 &Sal\\
\text{kickfb03}  & 100k	&-		& 2.0	&0.03	&0.002		&1  &2.5 &7 &Kr\\
\text{kickfb05}  & 100k	&-		& 2.0	&0.05	&0.002		&1  &2.5 &7 &Kr\\
\text{kickfb10}  & 100k	&-		& 2.0	&0.10	&0.002		&1  &2.5 &7 &Kr\\
\text{Zfb03}  & 100k	&-		& 1.0	&0.03	&0.001		&1  &2.5 &7 &Kr\\
\text{Zfb05}  & 100k	&-		& 1.0	&0.05	&0.001		&2  &2.5 &7 &Kr\\
\text{Zfb10}  & 100k	&-		& 1.0	&0.10	&0.001		&1  &2.5 &7 &Kr\\
\text{IMBHfb03}  & 100k	&200	& 1.0	&0.03	&0.002		&1  &2.5 &7 &Kr\\
\text{IMBHfb05}  & 100k	&200	& 1.0	&0.05	&0.002		&2  &2.5 &7 &Kr\\
\text{IMBHfb10}  & 100k	&200	& 1.0	&0.10	&0.002		&1  &2.5 &7 &Kr\\
\text{rh4fb05}   & 100k	&-		& 1.0	&0.05	&0.002		&1  &4 &7 &Kr\\
\text{rh4fb10}   & 100k	&-		& 1.0	&0.10	&0.002		&1  &4 &7 &Kr\\
\text{IMFfb05}   & 100k	&-		& 1.0	&0.05		&0.002		&1  &2.5 &7 &Sal \\
\text{IMFfb10}   & 100k	&-		& 1.0	&0.10   	&0.002		&1  &2.5 &7 &Sal \\
\text{W0}    & 100k	&-		& 1.0	&-   	&0.002		&1  &2.5 &3 &Kr \\
\text{W05}   & 100k	&-		& 1.0	&-   	&0.002		&1  &2.5 &5  &Kr \\
\hline
\text{can200k}	& 200k	&-		& 1.0	&-		&0.002		&1  &2.5 &7 &Kr\\
\text{fb10200k}	& 200k	&-		& 1.0	&0.10	&0.002		&1  &2.5 &7 &Kr\\
\text{IMBH200k}	& 200k	&400	& 1.0	&-		&0.002		&1  &2.5 &7 &Kr\\
\text{Z200k}	& 200k	&-		& 1.0	&-		&0.001		&1  &2.5 &7 &Kr\\
\text{kick200k}	& 200k	&-		& 2.0	&-		&0.002		&1  &2.5 &7 &Kr\\
\hline
\end{tabular}
\begin{tablenotes}
     \item[1] same as can100k, but with instantaneous stellar evolution
\end{tablenotes}
\end{threeparttable}
\end{table*}

\paragraph{}
Natal kicks, drawn from a Maxwellian distribution, are imparted to stellar remnants to account for asymmetry in core-collapse supernova explosions (see \cite{2019MNRAS.485.5752D} for further explanation). Larger natal kicks can contribute to the disruption of stellar binaries. However, this is not expected to impact the number of BHB mergers significantly, as most BHBs form dynamically. 

\paragraph{}
We initialize some models with a certain fraction of primordial binaries, $f$, such that there are $Nf/2$ primordial binaries, and $N(1-f)$ single stars. This is done by randomly selecting $Nf/2$ particles drawn from the IMF and assigning a companion based on the mass ratio distribution $f(q) = 0.6q^{-0.4}$ \citep{2007A&A...474...77K}. We predict that the primordial binary mass ratio distribution has little effect on the mass ratio of compact body mergers, as binaries are expected to undergo multiple exchange events prior to merger. The eccentricities, $0 \leqslant e \leqslant 1$, for primordial binaries are selected from a thermal (i.e., uniform) distribution \citep{1919MNRAS..79..408J}:

\begin{equation} \label{eq5}
    P(e) de = 2e de.
\end{equation}

 Finally the initial binary separations are drawn from a uniform distribution in $\log a_o$ (default \texttt{NBODY6} choice) , with a lower limit of $a_{min} = 0.1\text{AU}$, and an upper limit of $a_{max} = 10\text{AU}$ to produce both short and long period binaries. This choice ensures a majority of primordial binaries are not disrupted during early stellar evolution \citep{1975MNRAS.173..729H}.   

\paragraph{}
The simulations include tidal forces between the cluster and host galaxy. All models are placed in a circular orbit around a point-mass galaxy at a galactocentric distance of $23.3\text{kpc}$. Clusters are assumed to under-fill their tidal radius by a factor of three. Clusters are also initialized in dynamical equilibrium, assuming a Virial ratio of $Q = 0.5$ (absolute value of the ratio between gravitational potential energy and kinetic energy).

Most of the simulations are run up to a cluster age $t=12.5$ Gyr or terminated if the particle number reaches $0.3\text{N}$ due to tidal dissolution. \citet{2019MNRAS.485.5752D} include further details on the initial conditions, as well as on the cluster models used. 

\paragraph{}
We resort to an extended range of initial conditions to represent the wide range of sizes, masses, concentrations and metallicities of real-world mid sized GCs. While in principle the underlying physics of GCs (such as the natal kicks and primordial binary fraction) should remain the same irrespective of other initial conditions for real clusters, the actual values realised in nature for these degrees of freedom remain debated topics in the literature. Therefore, we explore a number of different models to investigate systems that differ in primordial binary fraction and natal kick distribution. See Section \ref{subsec:data_model} for a discussion on our model comparison and validation against LIGO results. For an investigation of how varying the binary fraction and natal kicks impact key results see Appendix \ref{appendix:split} instead.

\section{In-Cluster BHB Mergers}
\label{sec:in-cluster}
From our 58 simulations, there are 97 stellar BHB mergers within the clusters: 10 from the 11 50K simulations, 85 from the 42 100K simulations and seven from the five 200K simulations. These merger counts are qualitatively consistent with those reported by other studies with GCs of similar size and parameters \citep{2010MNRAS.402..371B}. For each merger, we record the merger time, component masses, mass ratio, and mass loss. We also track the history of the two components, including information about previous dynamical encounters and initial conditions. 

\begin{figure}[htt]
\begin{center}
\includegraphics[width =\columnwidth]{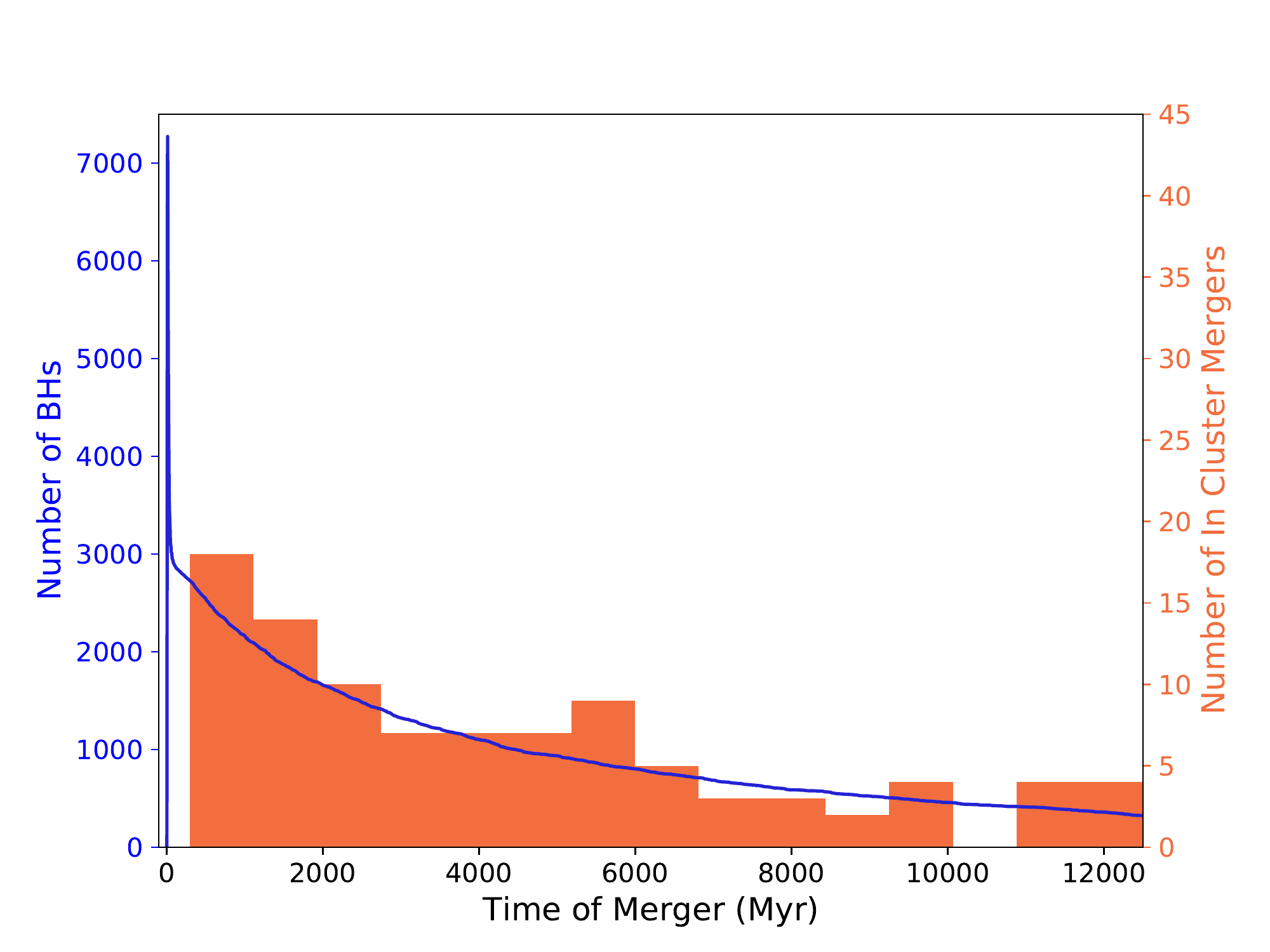}
\caption{In-cluster BHB merger rate (orange) and total number of BHs across all clusters (blue) as functions of time. As a cluster ages, the population of BHs evaporates, with the process driven primarily by dynamical ejections.}\label{Fig1}
\end{center}
\end{figure}

\paragraph{}
Figure \ref{Fig1} shows the evolution of merger number across all simulations as a function of the age of the star cluster. On average, mergers occur at a higher frequency earlier in the life of a cluster. This is mostly a result of the BH population evaporating due to dynamical ejection. As the number of BHs decreases, it becomes more difficult to form BHBs, and hence mergers become less frequent. We display the evolution of the number of BHs across all simulations as a function of time to show this population evaporation in Figure \ref{Fig1}. After the rapid formation of BHs by stellar evolution in the first few tens of Myr (with a substantial fraction ejected by natal kicks), we see a steady decrease over time as the cluster evolves and the BH number changes via dynamical processes (mergers and ejections). Interestingly, the number of mergers roughly follows this steady decrease in BH number (see Section \ref{subsec:incluster_ratio}). 

\paragraph{}
We define the quantity $N_{\text{retention}}$ as the number of BHs that remain in their host cluster after natal kicks. We also define $N_\text{merger}$ as the number of in-cluster BHB mergers that occur in a given simulation. Combining these two quantities we define scaled merger number:
\begin{equation}
    N_{\text{scaled}} = \frac{N_\text{merger}}{N_{\text{retention}}}.
\end{equation}

Investigating the relationship between primordial binary fraction, $f$, and $N_{\text{scaled}}$, we test for a non-parametric correlation using Spearman's rank-order correlation test\footnote{Note that for $f$ used in multiple simulations (eg, two simulations were run with $f = 0.05$), we simply use the total number of in cluster mergers and total number of retained BHs across these models.}. We find a correlation coefficient of $\rho = -0.54$ with a p-value of $0.27$, meaning that there is no statistically significant correlation between $f$ and scaled merger number. When testing for the correlation, we only use canonical simulations ($f = 0$) and simulations with primordial binaries, with all other initial conditions held constant. For example, we use the ``fb05" simulation in our correlation test, but not the ``IMBHfb05" simulation, as the presence of an IMBH may confound the results. The scaled merger number for $f = 0$ (canonical) simulations is higher than for any of the $f \neq 0$ simulations.

\subsection{Masses}
\label{subsec:incluster_mass}
\begin{figure}[ht]
\begin{center}
\includegraphics[width =\columnwidth, height = 25pc]{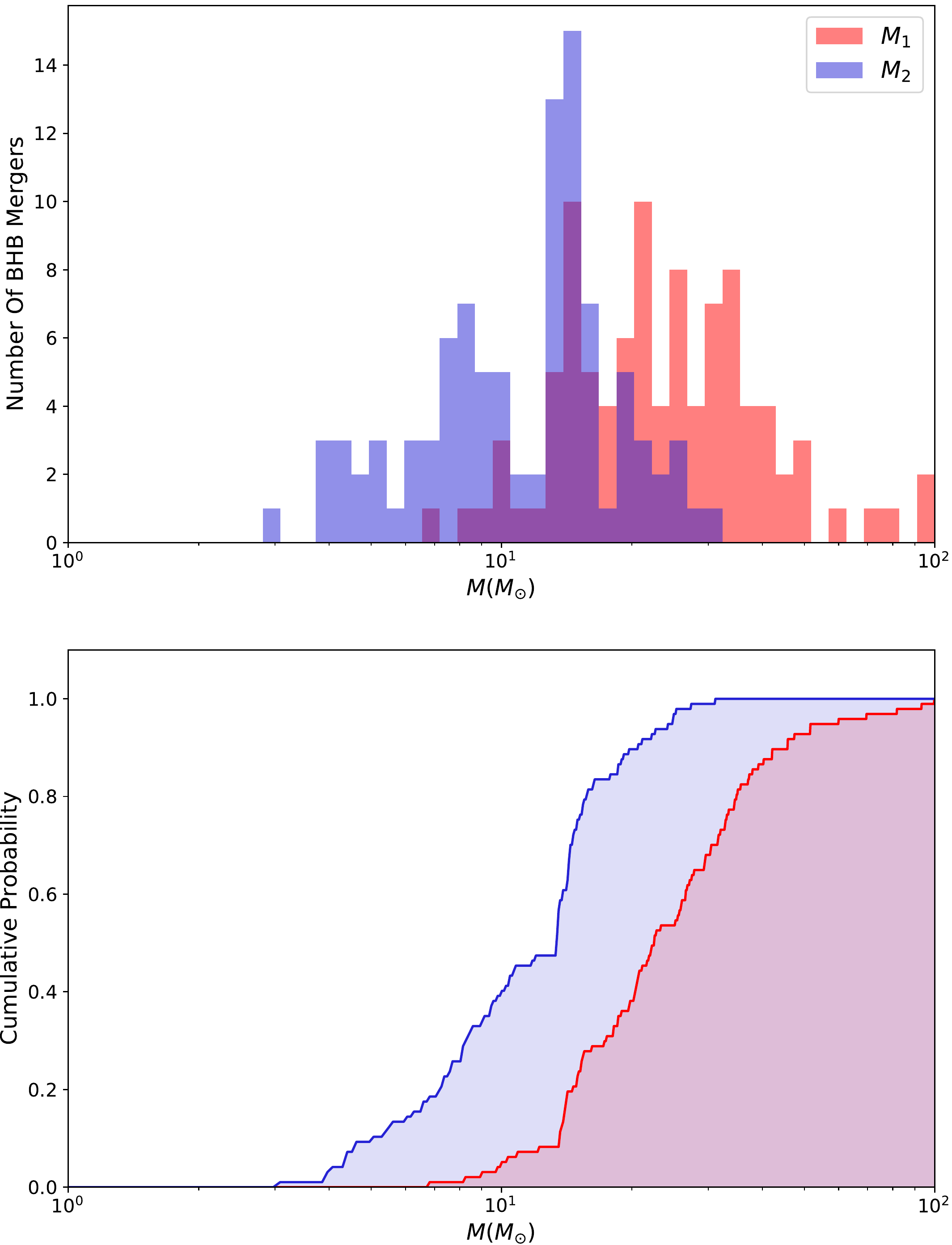}
\caption{Distribution of primary (red) and secondary (blue) masses for the population of BHBs that merge inside their host cluster, derived from all simulations in this study. The top panel shows the histograms for the component masses. The bottom panel shows the corresponding cumulative distribution functions (CDFs).}\label{Fig2}
\end{center}
\end{figure}

Two parameters of interest for the BHBs that merge within the cluster are the total binary mass and the mass ratio. We define the primary to be the heavier component, and the secondary to be the lighter companion. Figure \ref{Fig2} shows the primary and secondary mass distributions as a histogram (top panel), and a cumulative distribution (bottom panel). 

Both the primary and secondary mass distributions peak around $13 - 15 M_{\odot}$, with a mean mass of $24.9M_{\odot}$ and $14.5M_{\odot}$ respectively. The peak corresponds to the peak seen in the overall BH population distribution (which derives from the IMF and stellar evolution assumptions), as we verified by comparing the merging distributions to the mass distribution of a random sample of BHs across all models. It is therefore important to note that this peak is specific to the set of models being simulated, as the minimum progenitor mass is metallicity dependant.

Upon inspection of Figure \ref{Fig2}, the distribution of the primary BH masses appears to contain some high-end outliers. Indeed all five primary masses equal to or greater than $60M_{\odot}$ fall outside the range $\left[Q_1 - 1.5(Q_3 - Q_1), Q_3 + 1.5(Q_3-Q_1)\right]$\footnote{$Q_1$ and $Q_3$ are the lower and upper quartiles respectively.}, the standard convention for defining outliers \citep{James:2014:ISL:2517747}. Further investigation shows that each merger involves the product of the previous merger, i.e. the outliers occur when a BH merger product goes on to merge again multiple times. This chain of mergers is the only multi-merger string of its kind present in the simulations. All other BHs merge at most three times in their lifetime, whereas this chain represents seven total mergers. It allows a BH to reach masses above what is achievable through normal stellar evolution. We exclude the chain of seven mergers when comparing the in-cluster mass distributions to the ejected distributions (sections \ref{subsec:ejected_mass}). 

\paragraph{}
We do not compare Figure \ref{Fig2} to the output of similar studies in the literature, as the results are model and metallicity dependent. Many MC studies have utilized models with lower metallicities than the current paper \citep{2016PhRvD..93h4029R,2019arXiv190610260R, 2017MNRAS.469.4665P}. Expanding our simulation models to these lower metallicities is the subject of future research.

\subsection{Mass Ratios}
\label{subsec:incluster_ratio}
\begin{figure*}[httb]
\begin{center}
\includegraphics[width=\textwidth]{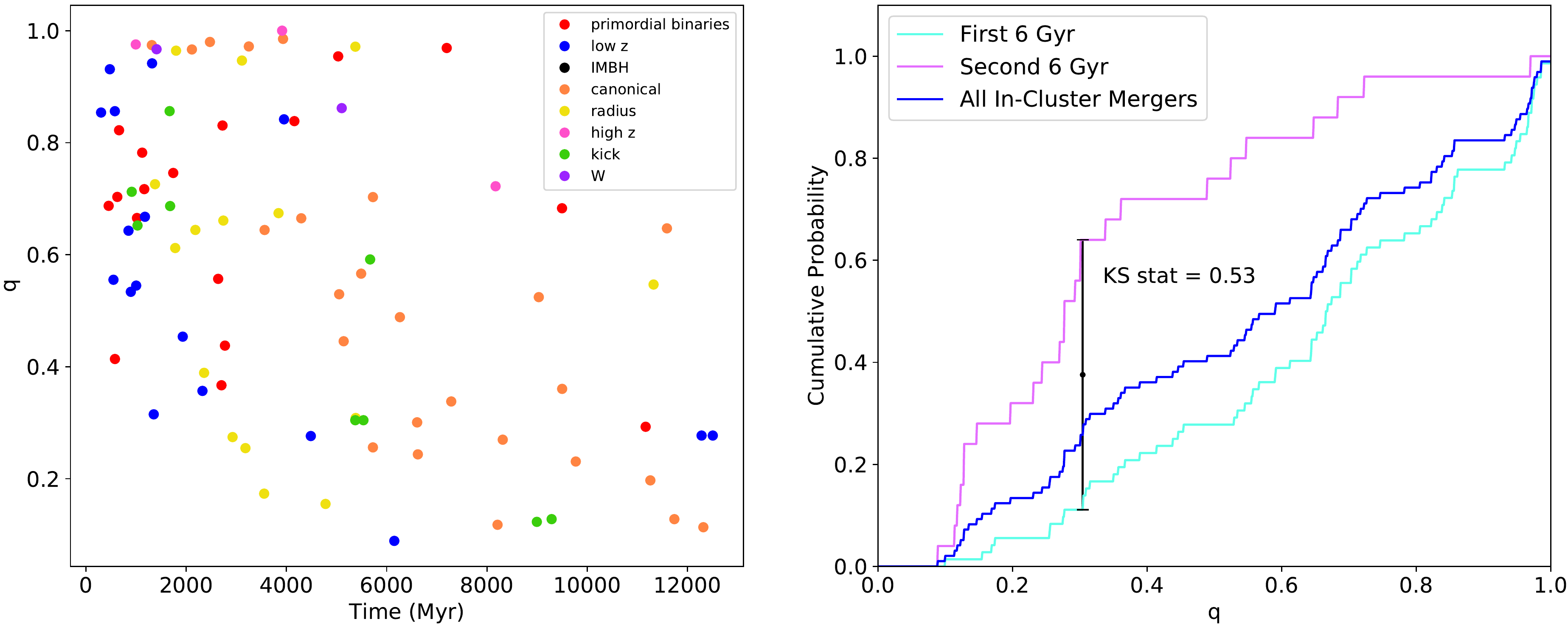}
\caption{Distribution of mass ratios for BHBs merging within their host cluster. The left panel shows the mass ratio $q$ versus time of merger (in Myr after initialization). We label mergers according to Table~\ref{tab:models}; primordial binaries (red), canonical (orange), low metallicity (blue), IMBH (black), large half-mass radius (yellow), high metallicity (pink), large natal kicks (green) and larger King concentrations (purple). The right panel shows the corresponding CDF for the mass ratios of the overall population (dark blue), for binaries that merge in the first 6 Gyr (cyan), and for binaries merging in the last 6 Gyr (pink). The KS statistic comparing the first 6 Gyr to the second 6 Gyr is 0.53, with a p value of $3\times10^{-5}$.}
 \label{Fig3}
\end{center}
\end{figure*}

We define the mass ratio as $q = M_2/M_1 \leqslant 1$. Figure \ref{Fig3} displays the cumulative $q$ distribution for merging BHBs (right panel), along with the evolution of $q$ with time (left panel). For $q \gtrsim 0.1$ the distribution is relatively flat. This result is somewhat surprising. Previous scattering experiments found that BHBs are likely to undergo multiple exchange events, where the lighter of the two binary components is preferentially ejected from the system in favour of the intruder \citep{1993Natur.364..423S}. This process serves to systematically increase $q$, as lower $q$ binaries exchange their relatively light component for masses closer to the primary. The distribution is therefore expected to skew towards higher mass ratios \citep{2016PhRvD..93h4029R,2019arXiv190610260R,2017MNRAS.469.4665P}. In the case of the first two studies, the majority of mergers are from ejected BHBs, whereas the third study only considers ejected binaries. We find the mass ratio distribution to vary significantly between in-cluster and ejected BHBs in this paper; see Section \ref{sec:in-cluster}.
\paragraph{}
Higher $q$ mergers tend to occur earlier in a cluster's lifetime. The left panel of Figure \ref{Fig3} displays a breakdown of merger $q$ by simulation type and merger time. All cluster models appear to follow the above trend roughly. We quantify the difference between early and late mergers by performing a Kolmogorov-Smirnov (KS) test, comparing the mass ratio distribution of mergers within the first 6 Myr to those in the second 6 Myr. With a KS statistic of 0.53 and a p-value of $3.03 \times 10^{-5}$, there is strong evidence against the null hypothesis that these two $q$ samples (first 6 Myr data and second 6 Myr data) come from the same underlying distribution.

The result in Figure \ref{Fig3} can be understood as a consequence of BH evaporation. There are no stellar-mass BHs present at the beginning of the simulations, but soon after initialization, stars above a mass of approximately $20M_{\odot}$ collapse \citep{2003ApJ...591..288H}, creating a population of remnant BHs. Once all massive stars have died, the BHs become the most massive bodies in the cluster and migrate to the cluster centre through two-body interactions as the system evolves toward partial energy equipartition \citep{2013MNRAS.435.3272T}. This leads to a high density of BHs in the core. Within this dense central population, compact bodies frequently undergo dynamical interactions, exchanging energy and often leading to the ejection of one or more bodies. These interactions and natal kicks reduce the BH population. As the number of BHs decreases, it becomes harder to form high $q$ binaries, because most BHs of similar mass have either already merged or have been ejected from the cluster.

\section{Ejected BHB Mergers}
\label{sec:ejected}
From the total of 58 simulations, there are 239 ejected stellar-mass BHBs: 43 from the 50K simulations, 162 from the 100K simulations and 35 from the 200K simulations. NBODY6 defines the ejection time for a binary as the time after cluster initialization when it the binary satisfies the Jacobi escape criterion; see \citet{2007MNRAS.377..465E} and \citet{1997MNRAS.284..811R}. We use Peters' equations \citep{1964PhRv..136.1224P} for the averaged orbital evolution to calculate the inspiral time of the BHBs upon ejection (equation \ref{eq1}). The merger time for escaped BHBs is the sum of the ejection time and this inspiral time. Here we make the simplifying assumption that all GCs originate at the same epoch, forming at around a redshift of 3.5 ($\sim$12 Gyrs ago), as this age roughly corresponds to the average age for Milky Way clusters. However, this assumption ignores the observed spread in cluster ages and the corresponding metallicity dependence \citep{2010MNRAS.404.1203F}. Using this method we find that a total of 20 ejected BHBs merge within the age of the Universe: two from the 50K simulations, 10 from the 100K simulations eight from the 200K simulations, resulting in a total merger number of 117 including the in-cluster mergers. The rest of the 239 do not merge within the age of the Universe, e.g. because $a_0$ is too large. For the rest of this paper, we refer to ejected BHBs that merge within the age of the Universe as merging escapers or merging systems. This prevalence for in-cluster over ejected mergers is also seen in N-body simulations of open clusters \citep{2017MNRAS.467..524B,2018MNRAS.473..909B,2018MNRAS.481.5123B}.

\paragraph{}
BHs that merge within the cluster can go on to form a new binary, so ejected BHBs can contain a BH which is the product of a previous merger. We refer to these as second-generation BHs. 
Merger chains systematically increase BH mass over generations but are unlikely to affect the orbital parameter distributions for the binaries. This is because all BHB undergo multiple exchange events and dynamical encounters where the orbital parameters are significantly altered, with the specific origin of the component BHs (whether through stellar death or merger) only altering the interaction cross section. We find no statistically significant difference between the orbital parameters of ejected BHBs with second-generation BHs and ejected BHBs with first-generation BHBs. 32 ejected BHBs contain at least one second-generation BHs, with four systems containing a third-generation BH. Only three merging systems contain a second-generation BH, and none contain a third-generation BH. In order to test if the observed difference is due to chance, we conduct a Pearson's Chi-squared test \citep{doi:10.1080/14786440009463897}, a statistical test used on categorical data to determine how likely the observed difference between data sets is purely due to chance. Here we have data belonging to two categories: whether or not an ejected system contains a second-generation BH, and whether or not a ejected system merges. With a p-value of 0.93, the results are consistent with the null hypothesis that there is no relationship between a system's ejection status (escape or merging escaper) and whether it contains a second (or higher) generation BH.

\subsection{Masses}
\label{subsec:ejected_mass}
\begin{figure}[hbt!]
\begin{center}
\includegraphics[width =\columnwidth, height = 25pc]{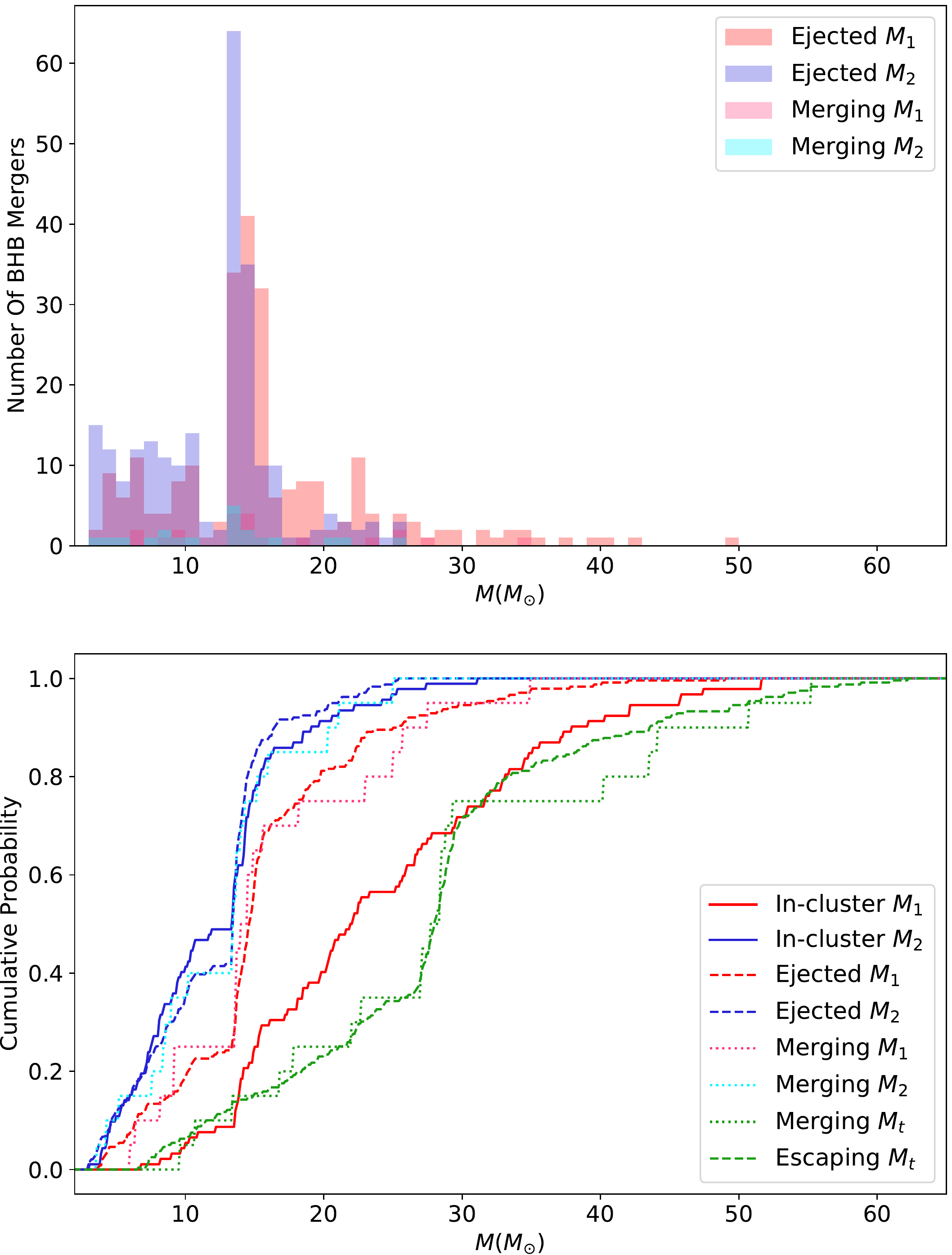}
\caption{Top panel: Distribution of BHB component masses: primary (red) and secondary (blue) masses of all ejected systems, and the primary (magenta) and secondary (cyan) masses for the subset of ejected systems that merge within the age of the Universe. Bottom panel: Corresponding mass CDFs for the different populations (dashed lines for the ejected systems, dotted lines for the subset that merge). We also include the CDFs for the in-cluster mergers for comparison (solid lines) and the total mass of the systems ($M_t = M_1 + M_2$) (green).}\label{Fig4}
\end{center}
\end{figure}

In Figure \ref{Fig4} we display the primary and secondary mass distributions for all ejected BHBs, and for the subset that are merging systems. Ejected BHBs exhibit a similar mass peak as in-cluster mergers (Section \ref{subsec:incluster_mass}). Note that we have excluded the outliers from the in-cluster $M_1$ data for comparison to the ejected $M_1$ distribution. Unlike their in-cluster counterparts, the ejected systems have no mass outliers. The distribution for $M_1$ in ejected systems is clearly skewed to lower masses when compared to the in-cluster $M_1$ distribution, and statistically different. In fact, a KS test yields a p-value of $7.1\times10^{-11}$ for these two distributions. The same cannot be said when comparing the corresponding $M_2$ distributions, with a p-value of 0.39. One possible explanation for the difference in in-cluster and ejected $M_1$ distributions is that $48\%$ of all in-cluster mergers involve a second or third generation BH (on average two or three times more massive than first generation products of stellar evolution), compared to only $15\%$ for ejected binaries.

\paragraph{}
The mass distributions for merging escapers closely match the ejected mass distributions. For ejected systems, there is no preference for merging based on either primary or secondary mass. This is somewhat surprising, as heavier systems emit more gravitational waves (equation \ref{eq1}) than lighter systems with the same orbital parameters. $\frac{dE_{GW}}{dt}/E_b$ is proportional to $M_1M_2(M_1 + M_2)$, meaning heavier systems merge more rapidly. However, it is more pertinent to instead consider the distribution of binary mass, $M_t = M_1 + M_2$. When considering $M_t$, the set of merging systems appears to be a representative sample of all ejected systems. The orbital parameters (eccentricity and semi-major axis) have a large impact on inspiral time: the integral in equation \ref{eq1} varies by several orders of magnitude depending on the initial eccentricity, and ejected systems display a large range in initial separations (see Section \ref{subsec:ejected_separation}), much larger than the range in BH and binary masses.

\begin{figure}[hbt!]
\begin{center}
\includegraphics[width =\columnwidth]{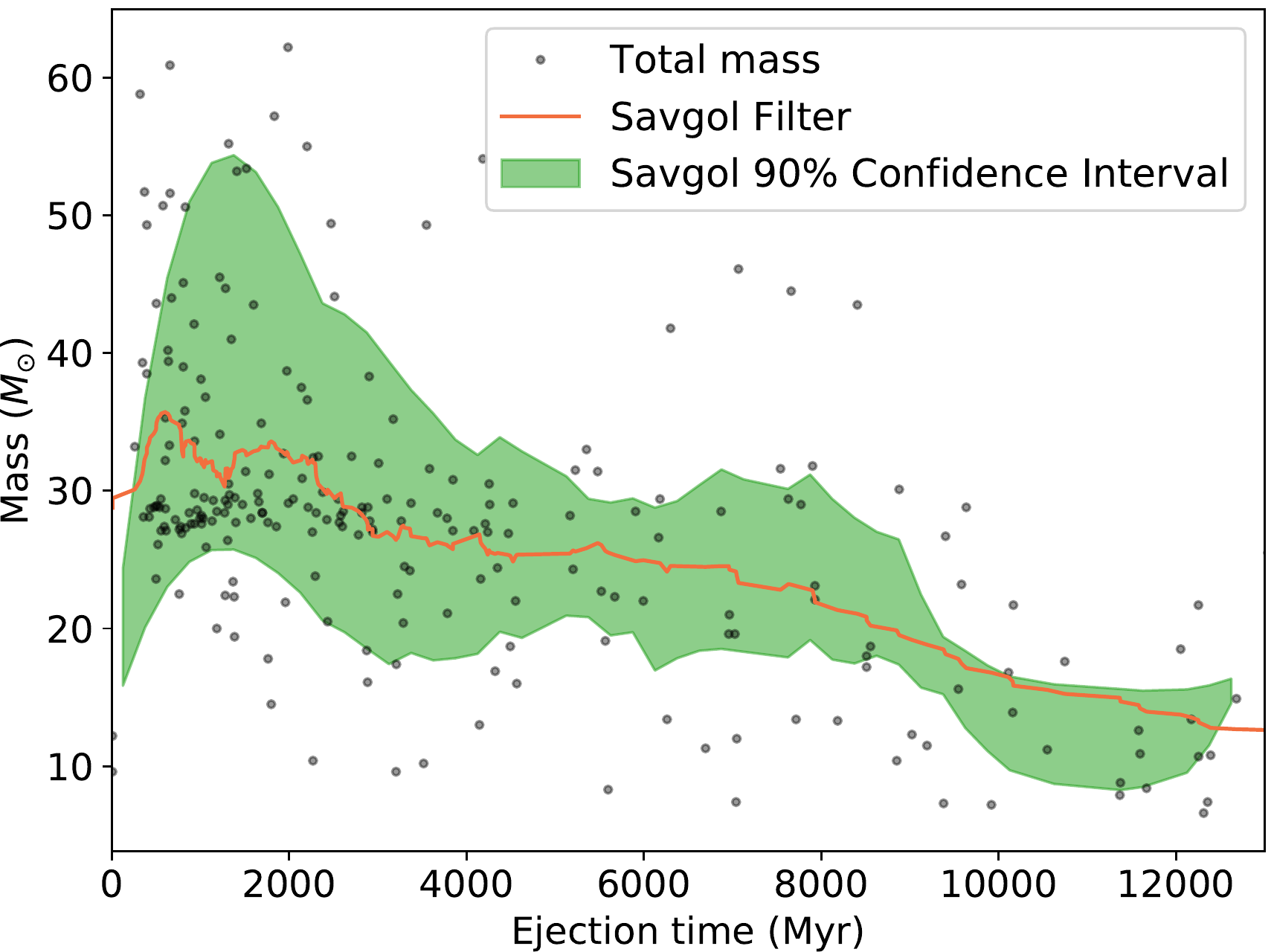}
\caption{Binary mass, $M_T$, of ejected BHBs as a function of ejection time. The grey points show each ejected BHB. The orange curve shows the trend in the data using a Savitzky-Golay filter. The symmetric 90\% confidence interval about the median (green band) is calculated by calculating the corresponding percentiles in 250Myr bins, smoothed through the Savitzky - Golay filter.}\label{Fig5}
\end{center}
\end{figure}

\paragraph{}
In Figure \ref{Fig5} we display the binary mass for all ejected BHBs as a function of ejection time. Although it appears that higher mass systems are ejected earlier, there is significant clustering around $28M_{\odot}$ at early times. We test for non-parametric correlation between binary mass and ejection time using Spearman's rank-order correlation test. We find a correlation coefficient of  $\rho = -0.52$ with a p-value of $10^{-17}$, indicating strong evidence against statistical independence ($\rho = 0$). That is, early high mass ejections are statistically significant.

\paragraph{}
A Savitzky-Golay filter \citep{1964AnaCh..36.1627S} is used to smooth the data to visualize this general tendency better. We also dipslay the symmetric 90\% confidence interval about the median. As the data is discrete, the confidence interval is calculated by first splitting the data into 250Myr bins and calculating the fifth and 95th percentile in each bin. We apply a Savitzky Golay filter on the set of percentiles to create smoothed $90\%$ confidence interval, symmetric about the median. These confidence intervals make it clear that higher mass binaries are ejected at earlier times. 

Similar trends have been observed in MC simulations of large GCs \citep{2016PhRvD..93h4029R}. This result is slightly counter-intuitive considering the ejection energy required for a binary to escape a cluster is proportional to the binary mass \citep{2003gmbp.book.....H}:

\begin{equation} \label{eq6}
    E_{ej} =\frac{GM_{GC}(M_1 + M_2)}{\sqrt{2^{2/3} - 1}R_h}, 
\end{equation}

\noindent where $R_h$ is the half mass radius, and $M_{GC}$ is the total mass of the cluster. It is therefore easier for lighter binaries to escape a given cluster, as $E_{ej}$ is lower. Superficially, this means binaries with higher-than-average masses are more easily retained by their host cluster, enabling more dynamic encounters, so that they are more likely to merge before ejection. However, we must consider the mechanisms which lead to ejection. Approximately $80\%$ of all ejected BHBs are ejected after a three-body encounter with another BH. Three-body scattering events harden binaries and can impart significant recoil velocities. Subsequent interactions can build up the centre-of-mass speed (through these recoils), possibly above the escape velocity of the cluster. The cross section for three-body interactions \citep{2018arXiv180711489C}, under the assumption that the binary is hard, can be approximated by 

\begin{equation} \label{eq7}
    \Sigma \approx \frac{2\pi G \left(M_1 + M_2 + m_3\right)a}{\sigma^2},
\end{equation}

\noindent where $M_1, M_2$ are the two binary component masses, $m_3$ is the third (single) body mass, $a$ is the binary semi-major axis, and $\sigma$ is the average stellar velocity within the cluster. With a larger cross section for more massive binaries, heavier BHBs have a higher interaction rate than their lighter counterparts, and hence experience a more rapid increase in velocity through successive interaction recoils. The rate of binding energy increase for hard binaries is approximated by

\begin{equation} \label{eq8}
    \frac{dE_b}{dt} \approx 2\pi \frac{G^2M_1M_2\rho}{\sigma}\epsilon,
\end{equation}

\noindent where $\rho$ is the mass density in the core, and $\epsilon$ is a scaling parameter that depends on the specifics of the interaction, with three-body scattering studies finding $\epsilon \sim 0.2 - 1$ \citep{1992MNRAS.259..115M,1996NewA....1...35Q}. Given $E_{ej} \propto (M_1 + M_2)$, and $ dE_b/dt \propto (M_1M_2)$, heavier binaries are expected to exceed their escape velocity earlier than their lighter counterparts. Heavier BHs are also produced earlier in a cluster's evolution due to shorter lifespans of high mass stars, and segregate to the dense core more rapidly, enabling quicker binary formation. The exception is second generation BHs, which form after the cluster has had time to host some mergers.

\subsection{Mass Ratios}
\label{subsec:ejected_ratio}
\begin{figure}[hbt!]
\begin{center}
\includegraphics[width =\columnwidth]{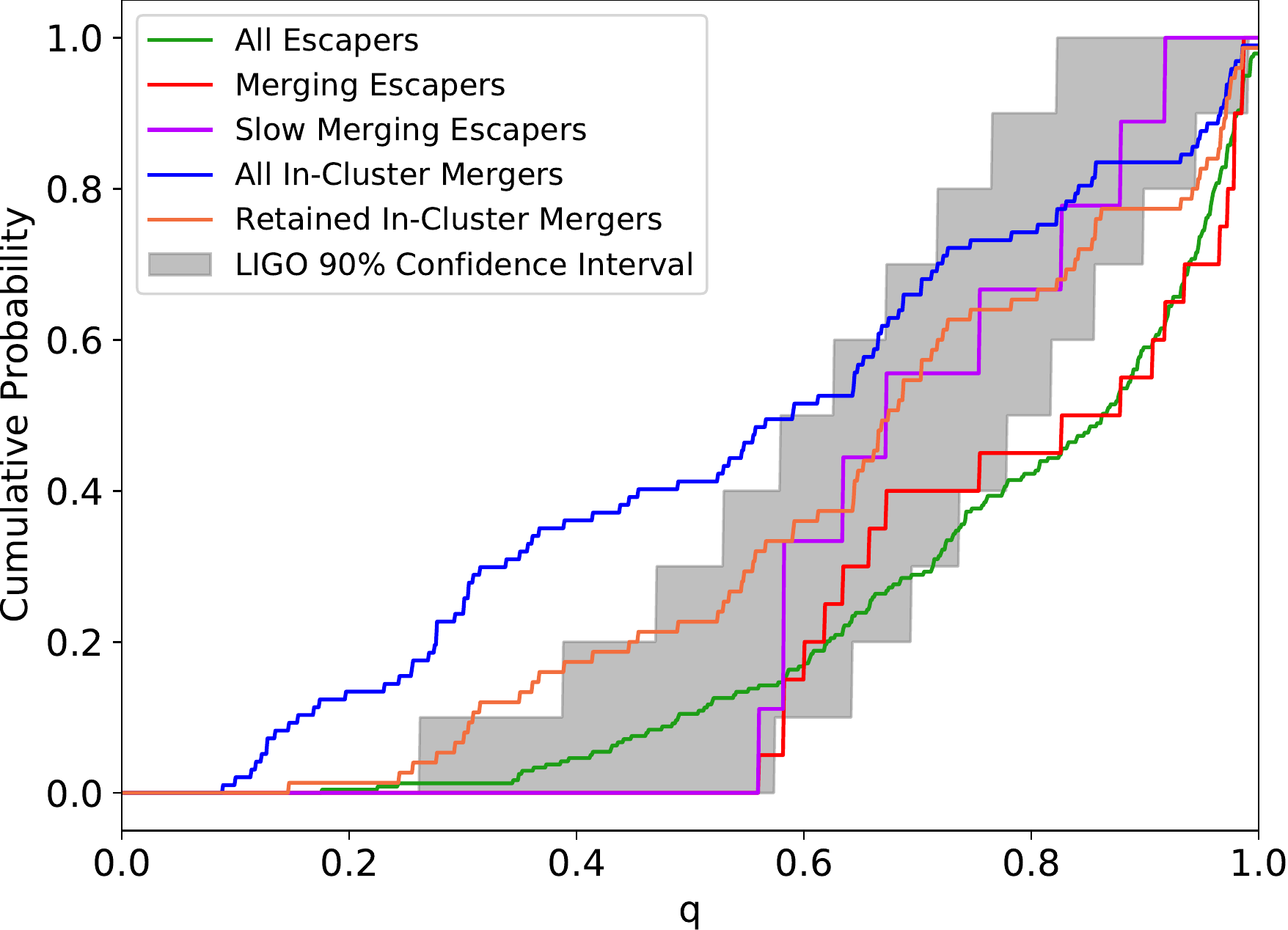}
\caption{Cumulative probability distribution for the mass ratios of BHB systems. The blue curve shows the distribution for all BHBs that merge within their host cluster. The orange curve shows the distribution in-cluster merging BHBs, excluding any BH which is flagged in post-processing as ejected after a previous merger through gravitational recoil (see section \ref{subsec:Gravitational_Recoil})}.  The green curve shows the distribution for all ejected BHBs, with the subset which merge within the age of the Universe displayed in red. Slow merging escapers (purple), are defined as ejected binaries that merge $ \geqslant 10^4 $ years after ejection. The distribution of inspiral times is bimodal, with one peak  $ \geqslant 10^4 $ years and the other peak $\leqslant 10^4$ years. The symmetric 90\% confidence interval for the mass ratio distribution given by the 10 LIGO events is displayed as a grey band for comparison.\label{Fig6}
\end{center}
\end{figure}

In Figure \ref{Fig6} we show the cumulative $q$ distributions for in-cluster mergers, ejected BHBs and the subset of ejected systems that merge within the age of the Universe. It is immediately clear that there is a preference for ejecting BHBs with larger mass ratios. Half of all ejected systems possess $q > 0.86$, because $q$ increases through the exchange of binary components (see Section \ref{subsec:incluster_ratio}). As a result, ejected systems have mass ratios much closer to unity than their in-cluster counterparts. 
The escaping BHBs and merging escaper distributions are closer to what has been found by \cite{2016PhRvD..93h4029R,2019arXiv190610260R} and \cite{2017MNRAS.469.4665P} than the flat distribution of in-cluster mergers. This has important implications when making inferences about formation channels from LIGO data. In Figure \ref{Fig6} we display the 90\% confidence interval for the CDF of the 10 LIGO mass ratios measured to date (see Section \ref{sec:ligo}). Taking into account uncertainty in the LIGO masses, there is a clear preference towards higher ratios, with probability $\geqslant 0.7$ for $q \geqslant 0.5$ for the 90\% confidence interval.

\paragraph{}
The ejected BHBs that merge display a similarly skewed $q$ distribution to the set of all ejected systems.  However, merging escapers have a low end cutoff (below which there are no BHBs in the population) at $q = 0.56$, compared to $q = 0.18$ for all ejected systems. There is no clear relationship between the mass-ratios, eccentricities or separations for ejected BHBs. The higher cutoff is predominately due to stronger production of gravitational radiation at higher mass ratios for a given binary mass, as $T_{\text{insp}} \propto \frac{1}{q}\frac{(1+q)^2}{M_T}$ is minimised when $q = 1$. 

\paragraph{}
The distribution of inspiral times for ejected mergers displays bimodality. In Figure \ref{Fig6} we display the set of merging escapers that have an inspiral time $> 10^4$ yr after ejection. We refer to these systems as slow merging escapers. Merging escapers with $ q \geqslant 0.9$ all merge $\leqslant 10^4$ years after ejection. Their slower counterparts appear to more closely match the LIGO 90\% confidence interval (see Section \ref{sec:ligo}). 

\paragraph{}
In Figure \ref{Fig6} we also display in orange the set of in-cluster mergers which exclude any system containing a second generation BH that would have been ejected through gravitational recoil (see section \ref{subsec:Gravitational_Recoil}). Compared to the full set of in-cluster mergers this retained population more closely matches the slow merging escapers, although there is still a significant number of mergers with $q < 0.5$ not seen in BH pairs merging after dynamical ejection.

\subsection{Eccentricities}
\label{subsec:ejected_eccentricities}
\begin{figure}[hbt!]
\begin{center}
\includegraphics[width =\columnwidth]{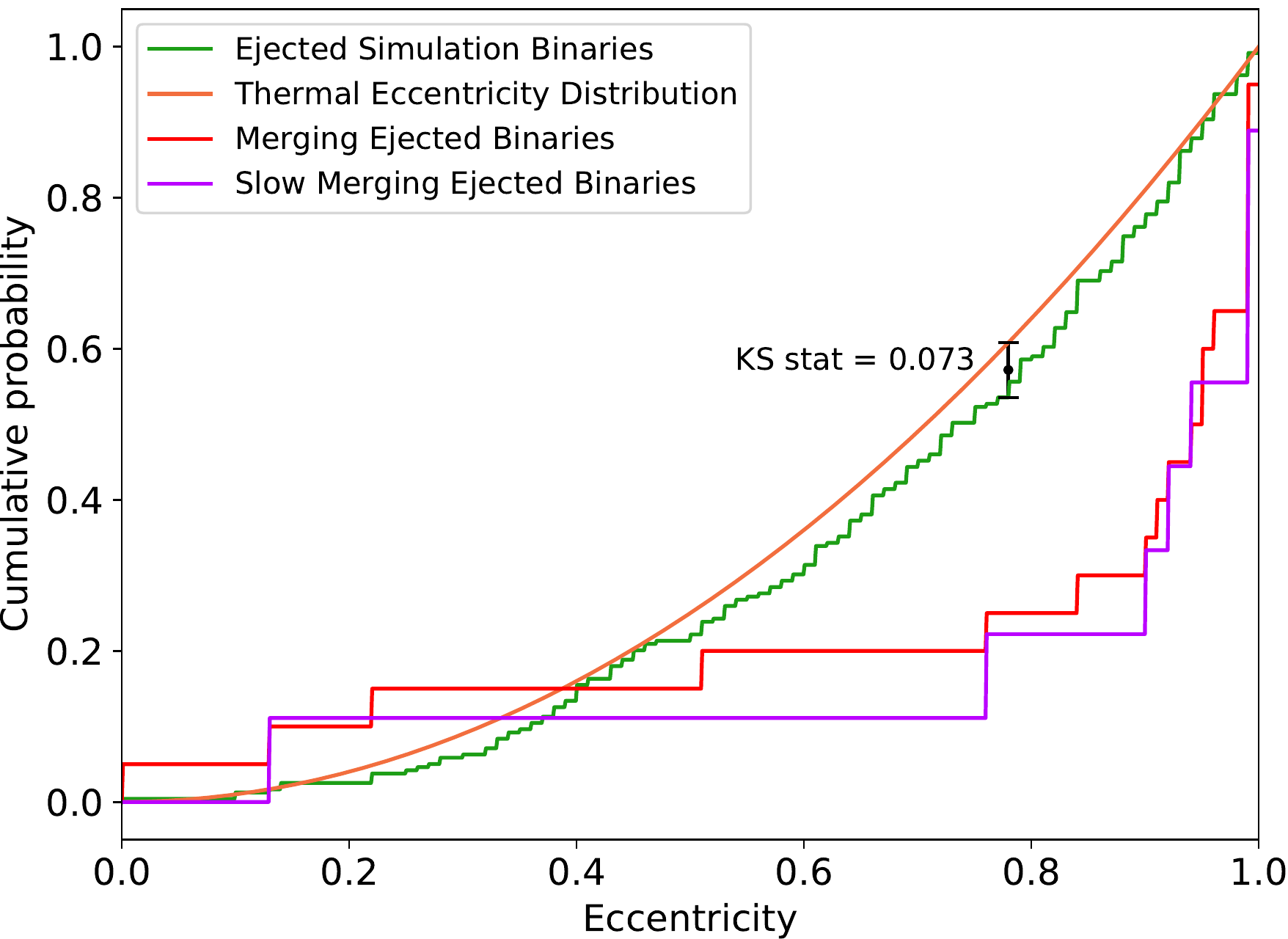}
\caption{CDF of eccentricities for different subsets of ejected BHBs. The eccentricities are taken at the moment when the binary is ejected from its host cluster. The green curve shows the distribution for all ejected systems, along with a theoretical thermal eccentricity distribution (orange) for comparison. Also shown is the ejected systems that merge within the age of the Universe (red) and the subset of these which which merge $ \geqslant 10^4 $ years after ejection. The KS statistic comparing the thermal distribution to all ejected systems is 0.073, with a p value of 0.15.}\label{Fig7}
\end{center}
\end{figure}

In Figure \ref{Fig7}, we show the cumulative distribution of eccentricities which ejected binary systems possess at the time of their escape from their host cluster. The distribution for all ejected binaries closely matches that of a thermal eccentricity distribution (equation \ref{eq5}), with a KS test giving a p-value of 0.15. In simulation models which include primordial binaries, the systems are generated with eccentricities drawn from a thermal distribution, but none of the ejected BHBs are primordial binaries; they form via exchanges, or from remnants which are never part of a primordial binary.  Although these initial eccentricities may still influence the distribution for binaries formed dynamically, 69\% of ejected BHBs come from clusters which did not begin with any primordial binaries. Hence the result suggests that ejected BHBs have had enough time to thermalize, by interacting with other particles/binaries and exchanging energy many times. 

We also display the cumulative distribution of eccentricities for merging ejected binaries and the slow subset. Higher $e$ systems tend to merge more frequently than ejected BHBs overall. Gravitational wave emission is stronger at higher eccentricities (equation \ref{eq1}), leading to faster inspiral. Ejected systems with high eccentricities progressively circularize after ejection due to gravitational radiation; all the ejected merging systems enter the LIGO frequency band with $ e \lesssim 10^{-4}$. This can also help explain why we do not see any noticeable preference for higher mass ejected binaries to merge, as eccentricity and separation (see Section \ref{subsec:ejected_separation}) affect the inspiral time more than mass.

\subsection{Separation}
\label{subsec:ejected_separation}
\begin{figure}[hbt!]
\begin{center}
\includegraphics[width =\columnwidth]{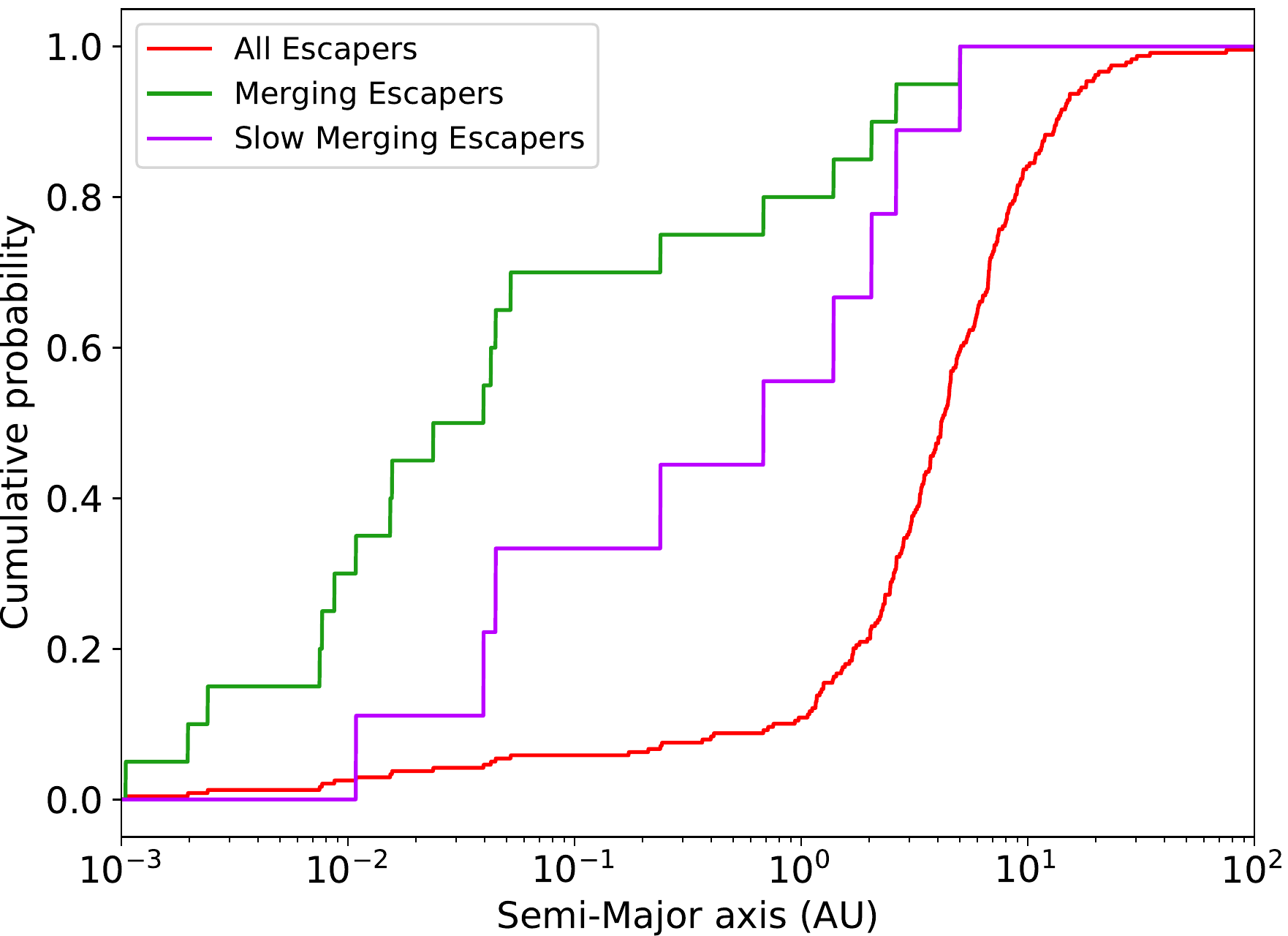}
\caption{CDF for the semi-major axis at the point of ejection on a log scale. The red curve shows the distribution for all ejected BHBs (red), with the merging escapers (green) and slow merging escapers (purple).}\label{Fig8}
\end{center}
\end{figure}

Figure \ref{Fig8} displays a log scale cumulative distribution for the semi-major axis of ejected BHBs. The green curve shows the distribution for the entire population of ejected binaries, the majority of which are ejected with $1 \text{AU} \leqslant a_0 \leqslant 10 \text{AU}$. The upper cutoff is partially a result of Heggie's law \citep{1975MNRAS.173..729H}, a statistical result which states that, on average, hard binaries become harder and soft binaries become softer in three-body encounters. A hard binary has $E_b > 1.2 \Bar{m} \sigma^2$, where $\sigma$ is the average stellar velocity and $\Bar{m}$ is the average stellar mass within the cluster. Most of the BHBs are ejected via velocity kicks imparted from subsequent strong three-body hardening. Soft binaries do not receive these hardening kicks and thus can never reach escape velocity, leading to binaries above a certain separation being too soft to be ever ejected. 

Binaries only have $a_0 \leqslant$1AU after multiple dynamical interactions (with the number dependent on the nature of the interactions), which in turn take time, leading to few systems with extremely low separations. Close binaries also lose energy to gravitational radiation and quickly merge within the cluster prior to ejection.

\paragraph{}
All merging ejected systems come from the low end of the overall $a_0$ distribution (red curve of Figure \ref{Fig8}, as these systems emit gravitational waves more strongly and inspiral more rapidly. For the slow subset that merge at least $10^4$ years after ejection, there appears to be a preference for larger separations, with all mergers above with $a_0 > 0.1$AU belonging to this subset, as to be expected based on arguments made above. These results, along with those from sections \ref{subsec:incluster_ratio} and \ref{subsec:ejected_eccentricities}, confirm that eccentricity and separation predominately affect whether or not a given ejected system merges within the age of the Universe. 

\paragraph{}
Although we attribute the prevalence of merging short period binaries to the low tail ($a \lesssim 1 AU$) of the separation distribution, these extremely short periods are somewhat surprising. To investigate if these short period binaries originate from a specific subset of initial conditions and/or simulations, we compare the separation data against all other variables (mass, escape time, eccentricity, mass ratio). Using a Spearman rank-order correlation test we find no correlation between variables, and find no discernible reason to conclude these short period binaries are in any way unique, leaving our starting hypothesis of the tail-of-the-distribution origin as the most plausible.

\paragraph{} Only 16\% of all mergers are from BHBs that are ejected from their host cluster. In contrast, similar studies using MC methods have found $\gtrsim 50\%$ of mergers from ejected systems \citep{2016PhRvD..93h4029R,2019arXiv190610260R,2017MNRAS.464L..36A}. It has been recently suggested that because the current state-of-the-art MC cluster codes, CMC \citep{2000ApJ...540..969J} and MOCCA \citep{2013MNRAS.429.1221H}, only incorporate strong interactions, the number of in-cluster mergers is being underestimated. \citet{2019arXiv190607189S} found that including weak encounters raises the proportion of in-cluster mergers. With weak encounters enabled, the total population of BHB mergers becomes dominated by in-cluster mergers, as we find in our simulations.

\paragraph{}
It is also possible that the initial particle size of our models may impact the number of ejected mergers, as previous work on smaller, less massive open clusters also find that in-cluster mergers dominate \citep{2017MNRAS.467..524B,2018MNRAS.473..909B}. These results imply that ejected mergers depend super-linearly on the mass of the clusters, as the proportion of ejected to in-cluster mergers appears to increase significantly with cluster mass. This highlights the importance of N-body modelling of smaller clusters such as those presented in this paper, as the majority of MC results focus on high mass clusters (M $\sim 10^5-10^6M_{\odot}$). Given that GCs roughly follow a $1/M^2$ mass distribution, clusters in our mass range better represent the bulk of the population of observed systems.

\paragraph{}
However, the initial density of the models could be a significant factor contributing to the lower proportion of ejected mergers in the current study. Binary separation upon ejection is primarily determined by the global properties of the cluster \citep{2000ApJ...528L..17P,2009ApJ...690.1370M}, as opposed to the properties of the binary. Physically this is because denser clusters have higher escape velocities, requiring binaries to harden more to allow for sufficient increase in speed through interaction recoil. Indeed it can be shown that for ejected systems one has \citep{2016PhRvD..93h4029R}

\begin{equation} \label{eq9}
    a_0 \propto \frac{M_{GC} \mu}{r_h},
\end{equation}

\noindent where $M_{GC}$ is the total cluster mass and $\mu = (M_1M_2)/(M_1 + M_2)$ is the reduced binary mass. Equation \ref{eq9} reinforces the point that smaller, more massive, and thus denser clusters eject tighter binaries, as they have higher escape velocities which allow the binaries to remain in the cluster and thus harden for longer. Our cluster models have similar initial $r_h$ to the models used in MC studies mentioned above \citep{2016PhRvD..93h4029R,2019arXiv190610260R,2017MNRAS.464L..36A}, but lower initial $N$, and hence lower $M_{GC}$. With models that have a higher initial value of $M_{GC}/r_h$, the mode of distribution of ejection separations, $P(a_0)$ (Figure \ref{Fig8}), decreases due to equation \ref{eq9}. In turn, this means that a greater proportion of all ejected binaries can emit sufficient gravitational radiation to merge in the age of the Universe, increasing the proportion of ejected versus in-cluster mergers.

\begin{figure}[h]
    \centering
    \includegraphics[width =\columnwidth]{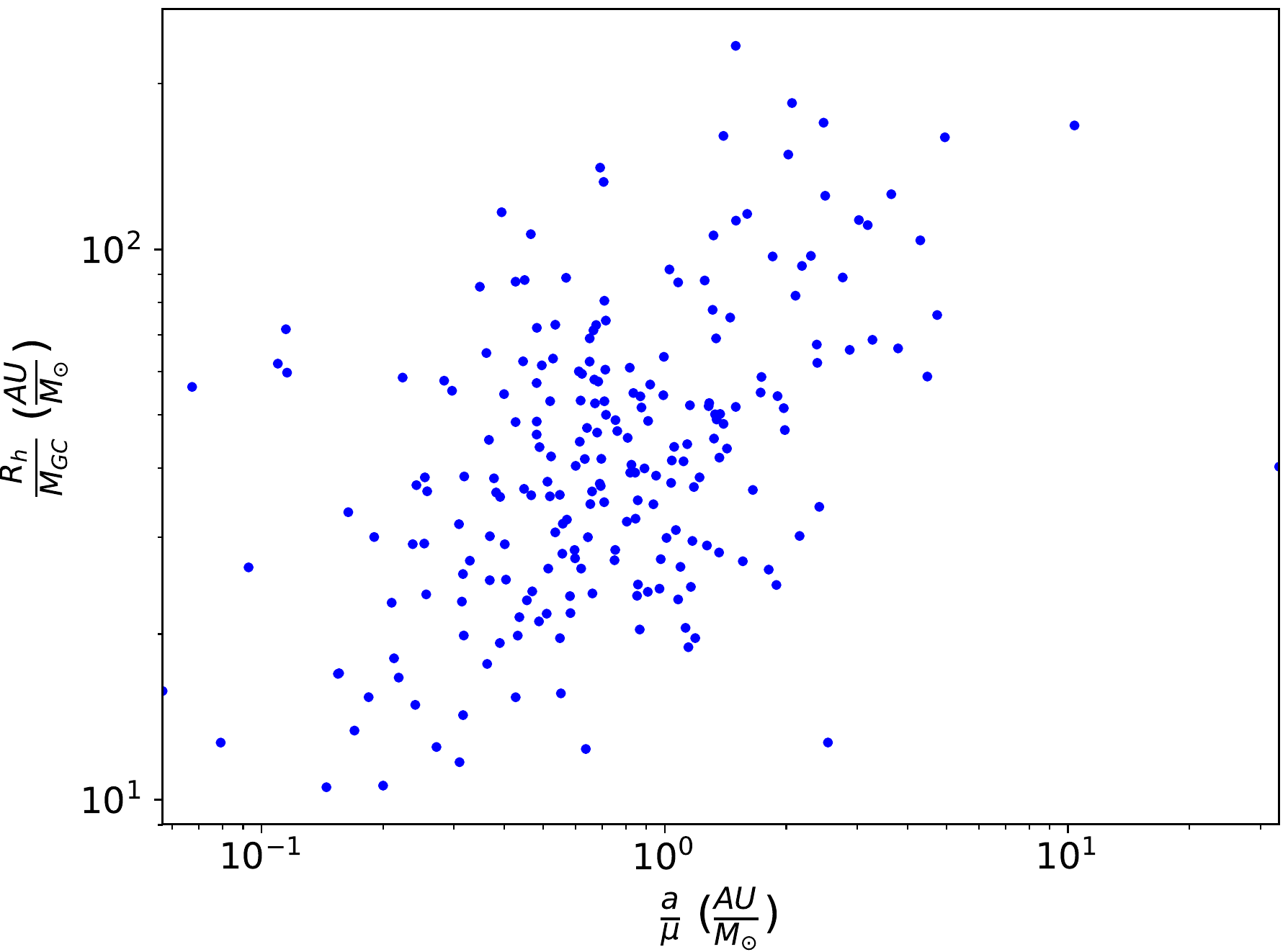}
    \caption{Scatter plot displaying the relationship between the ratio of the semi-major axis to the reduced mass for each ejected binary, $a/\mu$, and the ratio of the half-mass radius to cluster mass,$R_h/M_{GC}$, at the time each binary is ejected from the cluster. There is a clear positive correlation between the data.}
    \label{fig:scatter_plit}
\end{figure}

\begin{figure}[h]
    \centering
    \includegraphics[width =\columnwidth]{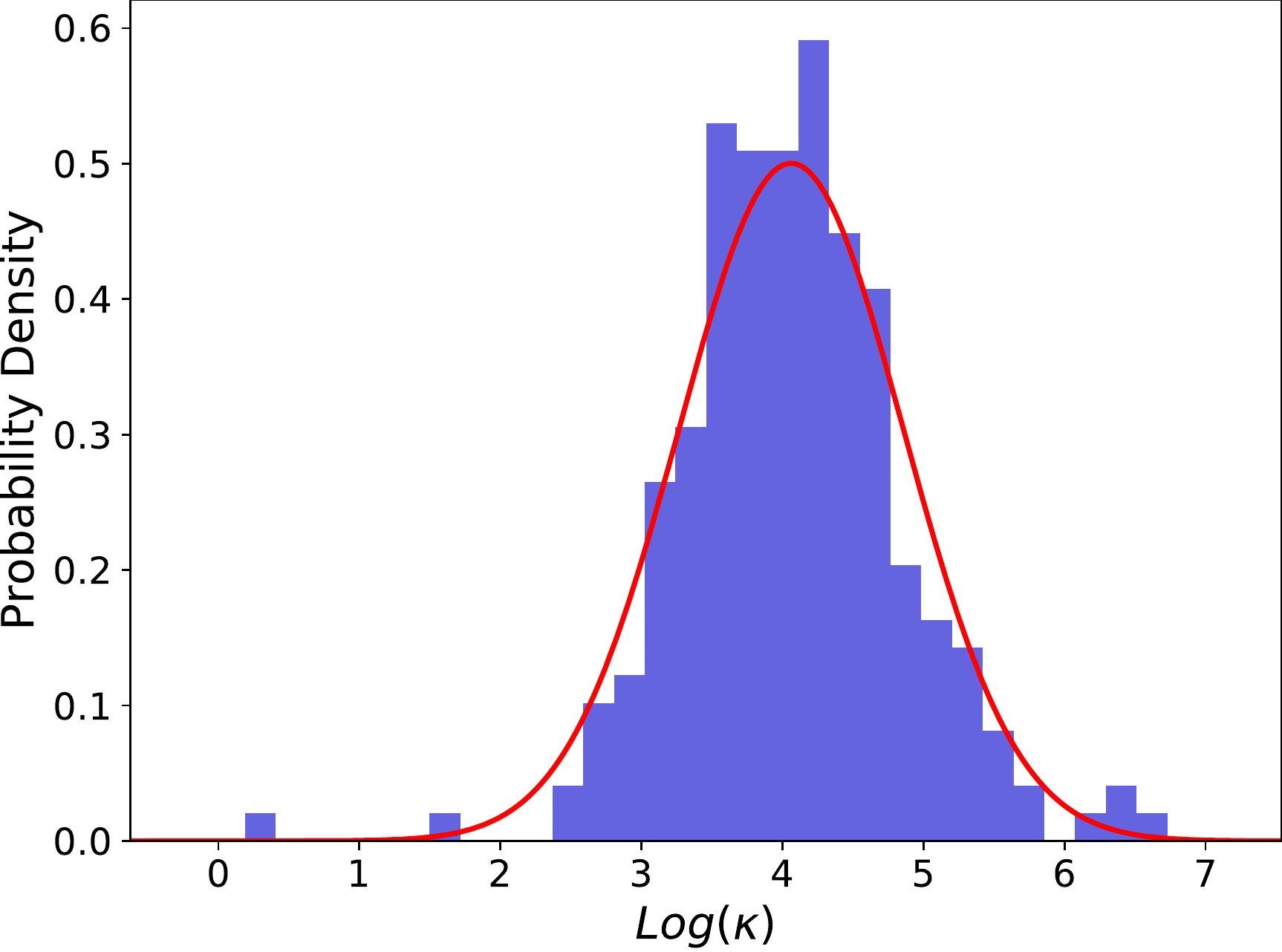}
    \caption{Probability density of $\log(\kappa)$ from all ejected binaries (blue). We fit a log-normal distribution to the data (red curve), finding a median value of $\sim 4$. }
    \label{fig:kappa_plot}
\end{figure}

We test the validity of equation \ref{eq9} to our data by plotting $a/\mu$ for every ejected binary against $R_h/M_{GC}$ at the time of ejection (Figure \ref{fig:scatter_plit}), finding, as predicted, a clear positive correlation. We quantify the non-parametric correlation between $a/\mu$ and $R_h/M_{GC}$ using Spearman's rank-order correlation test, finding a correlation coefficient $\rho = 0.43$ with a p-value of $10^{-11}$. This indicates strong evidence against statistical independence ($\rho = 0$). Thus our results are consistent with the prediction that denser clusters eject tighter binaries.

\paragraph{}
Following \citet{2016PhRvD..93h4029R}, we define

\begin{equation}
    \kappa = \left(\frac{R_h}{M_{GC}}\right) / \left(\frac{a_0}{\mu}\right)
\end{equation} 

In Figure \ref{fig:kappa_plot}, we show the distribution of $\kappa$ from all ejected binaries, plotted on a log scale. We fit a log-normal distribution to the data. Figure \ref{fig:kappa_plot} agrees strongly with the corresponding plot presented in \citet{2016PhRvD..93h4029R} (Figure 2, top panel), as both data roughly follow a log-normal distribution, with a median $\log(\kappa)$ value of $\sim 4$.

\section{LIGO Events}
\label{sec:ligo}
In this section we compare the results from sections \ref{subsec:incluster_ratio} and \ref{subsec:ejected_ratio} to the 10 BHB mergers observed and published by LIGO to date. We take the LIGO data from the catalogue of compact binary mergers observed by LIGO and Virgo during the first and second observing runs, O1 and O2 \citep{2018arXiv181112907T}. In order to incorporate the uncertainties in the LIGO masses, we employ a bootstrapping technique. We draw a random sample from the posterior distributions calculated using Bayesian inference \citep{2015JPhCS.610a2021V}. We use combined posterior distributions from two waveform models: an effective precessing spin model, IMRPhenomPv2 \citep{2014PhRvL.113o1101H,2016PhRvD..93d4007K} and a fully precessing model, SEOBNRv3 \citep{2014PhRvD..89h4006P,2014PhRvD..89f1502T}. Parameters are quoted in the source frame.

\paragraph{}
Although there are now a number of published BHB merger detections from LIGO's third observing run, we exclude these in our analysis as only one of these events is quoted as a BHB coalescence with reasonable confidence \citep{2020arXiv200408342T}. Moreover, the event mass posteriors required for our rigorous statistical analysis are not publicly available at the time of writing. We plan on including the set of O3 events in a future publication, once the full catalog is released publicly.

\subsection{Mass Ratio}
\label{subsec:ligo_ratio}
In Figure \ref{Fig6}, we display the simulated cumulative mass ratio distributions, overlaid with the LIGO data. The shaded region represents the $90\%$ confidence bounds for the LIGO distribution, symmetric about the median. We calculate these bounds by sampling from the LIGO primary and secondary mass posteriors. For each of the 10 events, we randomly pair an $M_1$ value drawn from the posterior with an $M_2$ value from the secondary mass posterior, without replacement, and this is repeated 10000 times. This process gives us a $q$ posterior for each detected BHB. Under the convention $M_1 \geqslant M_2$, these distributions have a sharp cutoff at $q=1$. We then randomly match each $q$ value from one distribution to one from each of the other nine event posteriors. In effect, we now have a set of 10000 $q$ samples used to construct CDFs, where each sample has 10 $q$ values, one from each event. The $90\%$ confidence intervals for the population distribution are computed by calculating these confidence intervals at each $q$ value. In practice, this means that the confidence bounds for the cumulative distributions contain points from multiple sample CDFs.

An alternative method is to calculate the fifth and 95th percentiles for the 10 events, using our generated $q$ posteriors. Two CDFs are then constructed, one from the 10 upper percentiles, and one from the 10 lower percentiles. The $90\%$ confidence interval band is the region bound by these two CDFs. However, for this analysis, we only consider the first method discussed.

\paragraph{}
As can be seen in Figure \ref{Fig6}, the LIGO data is skewed towards higher mass ratios. We pair each of the 10000 bootstrapped LIGO CDFs with the $q$ CDF of a given simulated population (such as the in-cluster mergers), and calculate the corresponding KS statistic (the maximum difference between the cumulative distributions). We then calculate the two-sided critical KS value for a given confidence level. This is the minimum KS statistic that is required to conclude that the differences between the two CDFs are statistically significant, i.e., the minimum distance to reject the null hypothesis. The critical value is defined by \citet{1987MNRAS.225..155F}

\begin{equation} \label{eq10}
    \text{KS}_{\text{crit}}(n,m,\alpha) \approx \text{K}_{\text{inv}}(\alpha)\sqrt{\frac{n+m}{nm}},
\end{equation}

\noindent where $\text{K}_{\text{inv}}$ is calculated from the inverse of the Kolmogorov distribution, $\alpha$ is the level of significance, and $n$ and $m$ are the number of data points being compared. For the sake of the current analysis, we set the level of significance at $\alpha = 0.05$. When sampling from the LIGO mass posteriors, we draw 10000 masses per distribution in order to sample the tails adequately. The proportion of KS statistics above this critical value indicates the proportion of the 10000 bootstrapped LIGO CDFs that are inconsistent with the null hypothesis, at 95\% confidence.

The above method only allows for comparison with our simulated populations, which we refer to as direct comparison. Alternatively, we wish to determine if our simulated dataset is robust enough to allow us to generalize our comparisons to the overall population of mergers being discussed (i.e., all merging escapers, not just the merging escapers present in our simulations). In effect, we want to know if our simulated data can be treated as a representative sample of the detectable population. We randomly pair each of the 10000 bootstrapped LIGO CDFs with CDFs constructed from samples drawn, with repetition, from our simulated $q$ distributions (in-cluster, ejected, merging escaper and slow merging escaper distributions). Because the critical KS value (equation \ref{eq10}) is a function of the size of the two distributions being compared, our new CDFs are constructed from samples which are the same size as the population from which the sample was drawn. For example, when comparing in-cluster mergers to observed events, we randomly draw, with repetition, 97 mass ratios from our distribution of 97 in-cluster mass ratios. In practice, this means these 10000 sampled distributions are not identical. We refer to this set of comparisons as sampling comparisons.  


\begin{table*}[t]
\begin{center}
\begin{tabular}{ |c|c|c|c|c| } 

 \hline
  \textbf{Simulated population} & \textbf{Size} & \textbf{Critical KS value} & \textbf{$p_{reject}$ [sampling]} & \textbf{$p_{reject}$ [direct]} \\ 
  \hline
 In-cluster  & 97 & 0.45 & 0.24 & 0.15\\ 
 Retained In-cluster & 75 & 0.46 & 0.02 & 0.007\\
 Ejected  & 239 & 0.44 & 0.49 & 0.48 \\ 
 Merging escapers & 20 & 0.53 & 0.30 & 0.07 \\
 Slow merging escapers & 9 & 0.62 & 0.13 & 0.01 \\

\hline
\end{tabular}
\end{center}
\caption{Comparisons of the mass ratio distributions between LIGO events and various populations of simulation BHBs. The simulation population being compared and its size is listed, along with the corresponding critical KS value at 95\% significance. We also list the probability that the null hypothesis is rejected at 95\% confidence, corresponding to the proportion of KS statistics above the critical value. Comparison is conducted both directly with our simulated CDFs $p_{reject}$ [direct], and with a set of CDFs constructed from samples drawn, with repetition, from the simulated distributions $p_{reject}$ [sampling]. The in-cluster population consists of all BHBs that merge within their host cluster, the retained population is the subset of in-cluster mergers that takes into account ejection of mergers remnants through gravitational recoil (estimated in post-processing), ejected systems are any BHBs that escape their host cluster, merging escapers are the subset of ejected systems that merge within the age of the Universe, slow systems are the subset that merge $\geqslant 10^4 \text{yr}$ after ejection. For all comparisons the size of the LIGO distribution is 10, corresponding to the 10 BHB mergers detected in O1 and O2.} \label{tab:ligo_comp}
\end{table*}

\paragraph{}
Table \ref{tab:ligo_comp} presents the mass ratio comparison between different simulated cluster populations and the LIGO data. We quote comparisons as the probability that the null hypothesis is rejected, at 95\% confidence. When considering in-cluster and merging escapers without gravitational recoil, the results when sampling from simulated data indicate that the LIGO mass ratios more closely match mergers that occur within the simulated GCs. These results are somewhat surprising; a cursory look at Figure \ref{Fig6} would seem to indicate the opposite. However, the probability of rejecting the null hypothesis (for sampling comparison) only differs by 0.06 between these two simulated populations. Indeed when compared directly to the simulation distributions, this result is flipped; the LIGO data matches the merging escapers better than in-cluster mergers. The mass ratio distribution is approximately flat for in-cluster mergers and skewed to higher ratios for ejected mergers. Thus, these findings offer a dynamical formation pathway for BHB mergers that agrees with the nearly flat/left skewed mass ratio distribution found by the LIGO Scientific Collaboration \citep{2018arXiv181112940T}. However, the fact that the favoured distribution is different between the direct and sampling comparisons, along with the corresponding differences in rejection probabilities, indicates that our simulated populations for in-cluster and merging escapers are not large enough to be treated as fully representative samples.

\paragraph{}
However, when we take into account gravitational recoil, the in-cluster mergers becomes a much better match to the LIGO data. Both the sampling and direct comparison for the retained in-cluster mergers have a higher rejection probability than any other population. This is unsurprising, as Fig.~\ref{Fig6} shows that accounting for gravitational recoil predominately reduces the number of low mass-ratio mergers (the majority of which contained a merger remnant BH), producing a similar low $q$ cutoff seen in the LIGO distribution. We note that excluding all in-cluster mergers containing a second-generation BH produces a mass-ratio distribution almost identical to the ejected population. This result is very  significant, as it implies that the LIGO data cannot be solely from common envelope evolution of field BHBs, as these should contain no second generation BHs. However, we again note the difference in rejection probabilities between sampling and direct comparisons as an indication that inference from the current sample of events observed is not currently robust enough to draw a firm conclusion.

\paragraph{}
For ejected systems, the slow merging escapers match LIGO data the best for both direct and sampling comparison, as expected from the comparison between CDFs in Figure \ref{Fig6}. This result has important implications if ejected systems dominate over in-cluster mergers. Ejection times are similar between all ejected mergers and the slow subset (the latter represents approximately half of the total). At design sensitivity, LIGO is limited to a canonical horizon of $\sim 1640$Mpc for BHB inspirals with $M_1 = M_2 = 30\text{M}_{\odot}$\citep{2018LRR....21....3A}, corresponding to a redshift of 0.39. Assuming that clusters form $\sim 12$Gyr ago, then the canonical horizon distance corresponds to binaries that merge $\geqslant 6$ Gyr after star cluster formation. Slow merging escapers are thus more likely to be detected than escapers that merge quickly after ejection. As with the other simulated populations, we again point out the difference in direct and sampling rejection probabilities (table \ref{tab:ligo_comp}) as a caveat that a larger set of N-body runs would be highly beneficial to draw stronger inference. However, the ejected population appears to be sufficiently robust, with the rejection probability changing by 0.01 between sampling and direct comparison. Expanding our data-sets to achieve large sample simulated populations to allow for effective comparison will be the subject of future work. We likewise leave a full analysis of the mass dependence  \citep{2012arXiv1203.2674T}, and peak GW frequency \citep{2016PhRvD..93k2004M} to future studies. 

\subsection{Data-model Comparison}
\label{subsec:data_model}
It is important to consider how representative our simulations are of actual star clusters that host mergers detectable by LIGO. The models used are not an accurate sample of the real-world GC population in the local Universe. Making such a sample is intrinsically difficult as the underlying distributions for GC parameters are not entirely known, being limited by our ability to observe the clusters. Most of the information pertaining to these distributions come from Milky Way GCs, as extra-galactic clusters are in general too distant. There is no reason to think that these local GCs are representative of all GCs.

\begin{figure*}[hbt!]
\begin{center}
\includegraphics[width = \textwidth]{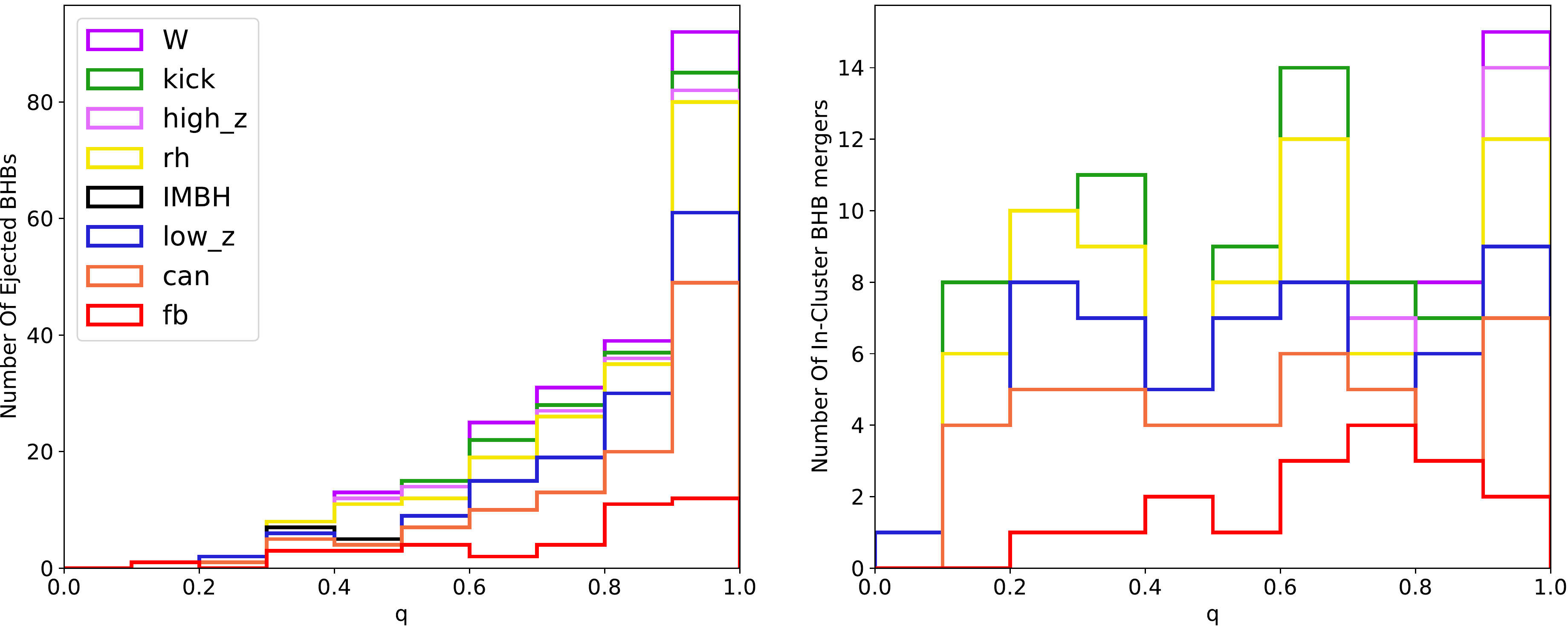}
\caption{Breakdown of the number of mergers in each cluster model per mass ratio; primordial binaries (red), canonical (orange), low metallicity (blue), IMBH (black), large half-mass radius (yellow), high metallicity (pink), large natal kicks (green) and larger King concentrations (purple). The left panel displays the mass ratio distribution for all ejected systems, and the right panel displays the mass ratio distribution for all in-cluster mergers.}\label{Fig9}
\end{center}
\end{figure*}

Instead, we rely on the relatively wide range of initial conditions to explore the parameter space. When making comparisons to LIGO observations, we must be careful to ensure that the trends we observe are not artefacts of our initial conditions. Figure \ref{Fig9} displays histograms of the in-cluster (right panel) and ejected (left panel) $q$ distributions, separated into each of the eight main models. The overall trends appear to be followed reasonably well by most models. The exception is the "high\_z" and "W" models for the in-cluster mergers, all of which have $q > 0.7$. If these two $q$ distributions were indeed flat, the probability of having five independent mergers above 0.7 is 0.0024. However, as these five mergers only represent $\sim 5\%$ of in-cluster mergers, their effect on the overall distribution is small, and are thus unlikely to significantly confound our results. The same can be said for the presence of two ejected BHBs originating from an IMBH model (black section in Figure \ref{Fig9}), both with $q < 0.45$.

For merging escapers and slow merging escapers, we display the initial condition information in table \ref{tab:model_breakdown}. The slow merging escapers appear to be a representative sample of all merging escapers regarding the proportion of mergers in each model. We also see the relatively large number of "can" and "low Z" ejected BHBs in the merging systems, likely due to the large number of simulations using these models. Only one merging escaper comes from an "rh" model (5\%), whereas 45 ejected systems  (18\%) come from this model. This result is because "rh" models have a larger initial $r_h$ than other models, leading to wider ejected binaries (equation \ref{eq9}). Overall, all these checks do not highlight the presence of significant bias in the analysis introduced by the choice of initial conditions.

\begin{table}
\begin{center}
\begin{tabular}{ |c|c|c|} 
 \hline
 &  \multicolumn{2}{|c|}{\textbf{Ejected population}}\\ 
  \hline
   \textbf{ Model} & \textbf{Merging} & \textbf{Slow merging} \\
  \hline
 fb  & 2 & 0 \\ 
 can  & 6 & 3 \\ 
 low Z & 5 & 3 \\
 IMBH & 0 & 0 \\
 rh & 1 & 0 \\
 high Z & 1 & 1\\
 kick & 4 & 2 \\
 W & 1 & 0 \\
\hline
\end{tabular}
\end{center}
\caption{Breakdown of the number of mergers in each cluster model, presented for merging escapers and slow merging escapers.} \label{tab:model_breakdown}
\end{table}

\section{CONCLUSIONS}
\label{sec:conclusion}
In this work we analyze the mergers of BHBs formed within simulated GCs, comparing the binary parameters of systems which merge inside their host cluster, to systems which merge after being ejected. The analysis is based on a novel set of direct NBODY6 simulations of realistic cluster models spanning a wide range of initial conditions, partially presented previously by \cite{2019MNRAS.485.5752D} in the context of structural GC properties. The simulations include single and binary stellar evolution, galactic tides, gravitational radiation, and other relativistic effects, making them an ideal tool to explore BHB mergers and compare with recent LIGO results. The key results are:

\begin{itemize}
    
    \item Cluster BH populations evaporate over time through merger and ejection, leading to in-cluster merger frequency decreasing as the GCs age. The number of BHs and the number of BHB mergers are strongly correlated over time with each other. \\ 
    
    \item We find no correlation between primordial binary fraction and scaled merger number. This indicates that primordial binaries have little impact on cluster BHB merger rates, likely because merging BHBs are dynamically formed. \\
    
    \item The in-cluster and ejected mass distributions are relatively similar, except that in-cluster merging BHBs tend to have slightly higher primary masses, as second generation BHs merge again in the cluster core. Both populations peak in their mass distributions per the overall BH population, resulting from the progenitor cutoff for forming BHs. Higher mass systems tend to be ejected earlier in a cluster's lifetime, because the three-body interaction rate is proportional to binary mass, enabling quicker ejection through successive interaction recoils. \\
    
    \item The eccentricity distribution for escaping BHBs closely matches a thermal distribution, indicative of multiple three-body interactions. Although primordial binaries (when present) are initialized with eccentricities drawn from a thermal distribution, none of the ejected BHBs are primordial. The subset of ejected BHBs that merge within the age of the Universe have an eccentricity distribution skewed closer to $e = 1$ when compared to the thermal distribution, a result of the stronger gravitational radiation production at higher eccentricities.  \\
    
    \item Only 8\% of all ejected BHBs merge within the age of the Universe, corresponding to approximately 16\% of mergers when including in-cluster mergers. The ratio of ejected to in-cluster mergers is significantly lower than in previous studies \citep{2016PhRvD..93h4029R,2019arXiv190610260R}. We attribute this discrepancy to the relatively high stellar densities in the GCs simulated by \cite{2019MNRAS.485.5752D}. \\
    
    \item The majority of escaping BHBs are ejected from their host cluster with separations between 1 and 10 AU. However, only systems with separations $\lesssim$ 1 AU produce sufficient gravitational radiation to merge within the age of the Universe. \\
    
    \item Ejected systems have a mass ratio distribution skewed towards unity, whereas in-cluster systems have an almost flat distribution for $q \gtrsim 0.1$. In addition, we observe a bimodality in the inspiral times after escaping the cluster, whereby all extreme mass ratio systems ($q > 0.95$) merge within $10^4$ years after ejection. \\
    
    \item We compare our simulated mass ratio distributions to the 10 LIGO BHB mergers detected in O1 and O2. Both slow-merging escaping BH pairs and in-cluster mergers that we estimate are not impacted by gravitational recoil match the LIGO O2 data reasonably well. However, the results are affected by low-number uncertainty; therefore, larger samples of observed mergers and a more extensive set of simulations are needed for robust confirmation of this tentative finding and to discriminate between the two possible populations. \\
    
    \item Finally we separate our simulations into sub-types based on the initial conditions. We find that the overall trends seen in the mass ratio distributions are mostly also seen in the individual simulation sub-types. We conclude that our selection of initial conditions does not significantly bias the mass ratio distributions. 
\end{itemize}

In a future extension, we plan to run models with lower metallicities to allow for a more comprehensive comparison with similar studies \citep{2016PhRvD..93h4029R,2019arXiv190610260R,2017MNRAS.469.4665P}, and to conduct a more systematic investigation into the distribution of real-world GC properties, using the results to expand to a more extensive and realistic simulation database. A full statistical analysis with LIGO O3 Data will also be conducted, incorporating horizon distance calculations to only include simulated BHB's that LIGO can detect. Finally, we will extend our models to different formation epochs, using a cosmological model of GC formation, and incorporate new prescriptions for gravitational recoil in NBODY codes through numerical relativity.

\section{ACKNOWLEDGEMENTS}

We would like to thank Ruggero De Vita, Ilya Mandel and Marcel Hohmann for useful discussions. Parts of this research are supported by the Australian Re-search Council Centre of Excellence for Gravitational Wave Discovery (OzGrav) (project number CE170100004). This research has made use of data, software and/or web tools obtained from the Gravitational Wave Open Science Center (https://www.gw-openscience.org), a service of LIGO Laboratory, the LIGO Scientific Collaboration and the Virgo Collaboration. LIGO is funded by the U.S. National Science Foundation. Virgo is funded by the French Centre National de Recherche Scientifique (CNRS), the Italian Istituto Nazionale della Fisica Nucleare (INFN) and the Dutch Nikhef, with contributions by Polish and Hungarian institutes. Numerical simulations have been performed on HPC clusters at the University of Melbourne  (Spartan) and at the Swinburne University (OzSTAR).

\section{CONFLICTS OF INTEREST}

None.

\appendix

\section{Primordial binaries and natal kicks}
\label{appendix:split}
As discussed in sections \ref{subsec:models} and \ref{subsec:data_model}, our models encompass a wide range of initial conditions, reflecting both the variety of masses, concentrations, and ages in the GC population of galaxies, and uncertainties in the input physical ingredients (natal kick distribution and primordial binary fraction). In nature the latter physical ingredients do not vary from one GC to the next, so it is important to assess if our modeling choices introduce biases in our results. To address this we split some of our main results by natal kick distribution and primordial binary fraction. 

\paragraph{}
Although we run systems with a number of different primordial binary fractions, due to the small sample size, we only compare clusters without primordial binaries to those with primordial binaries, instead of separating results into the specific primordial binary fraction. For natal kicks we compare clusters with $\sigma_k/\sigma_*$ = 1 to those with $\sigma_k/\sigma_*$ = 2. The key results we compare are in-cluster and ejected masses and mass ratios, along with ejected eccentricities and separations.

\begin{figure}[h]
    \centering
    \includegraphics[width = \columnwidth]{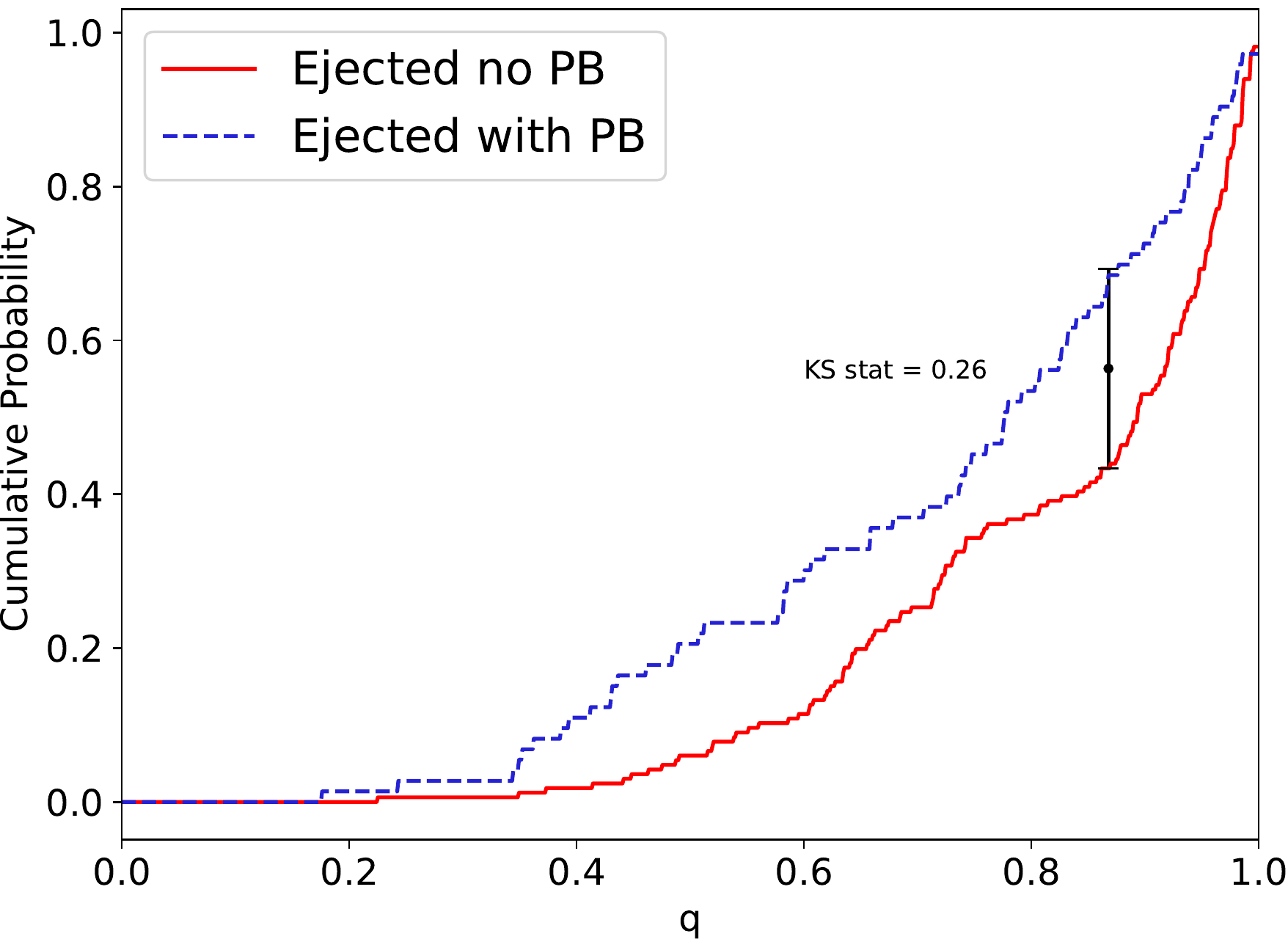}
    \caption{CDF of the mass ratios for ejected BHBs. The red curve shows the distribution for models without primordial binaries, and the blue curve shows the distribution for models with primordial binaries. The KS statistic is ~0.26, with a p-value < 0.01.}
    \label{fig:q_bin_split}
\end{figure}

\paragraph{}
 When comparing the models with and without primordial binaries using the KS test, only the separation and ejected mass ratio distributions yield strong evidence against the null hypothesis at 95\% confidence. Figure \ref{fig:q_bin_split} shows the ejected BHB mass ratios CDFs for models with and without primordial binaries. Although the difference between these sets of models is statistically significant, they still display the same prevalence of higher mass ratios. This suggests that while the effect of our choice of primordial binary fractions cannot be ruled out, its impact is expected to be modest. 

\begin{figure}[h]
    \centering
    \includegraphics[width = \columnwidth]{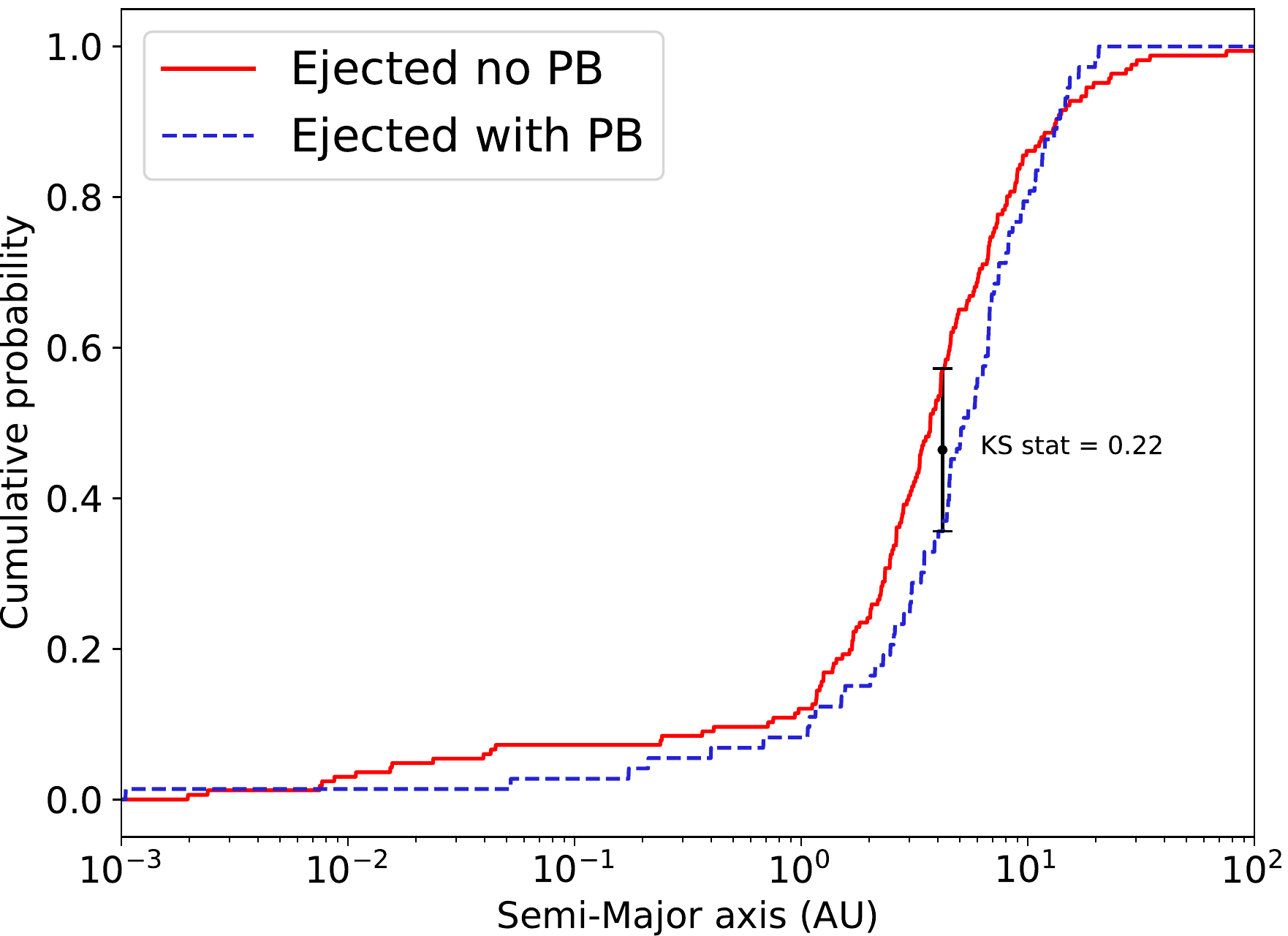}
    \caption{CDF of the semi-major axis at the point of ejection, plotted on a log scale. The red curve shows the distribution for models without primordial binaries, and the blue curve shows the distribution for models with primordial binaries. The KS statistic is ~0.22, with a p-value of 0.01.}
    \label{fig:sep_bin_split}
\end{figure}

Figure \ref{fig:sep_bin_split} shows the ejected BHB separation CDFs for models with and without primordial binaries. Using the KS test to compare ejected separations for the two different kick models, we find evidence against the null hypothesis at 95\% confidence. Similarly to the mass ratios discuss above, both distributions still display similar trends despite statistically significant differences. All other comparisons of ejected BHB separations for different initial conditions are consistent with the null hypothesis instead. 

\begin{figure}[h]
    \centering
    \includegraphics[width = \columnwidth]{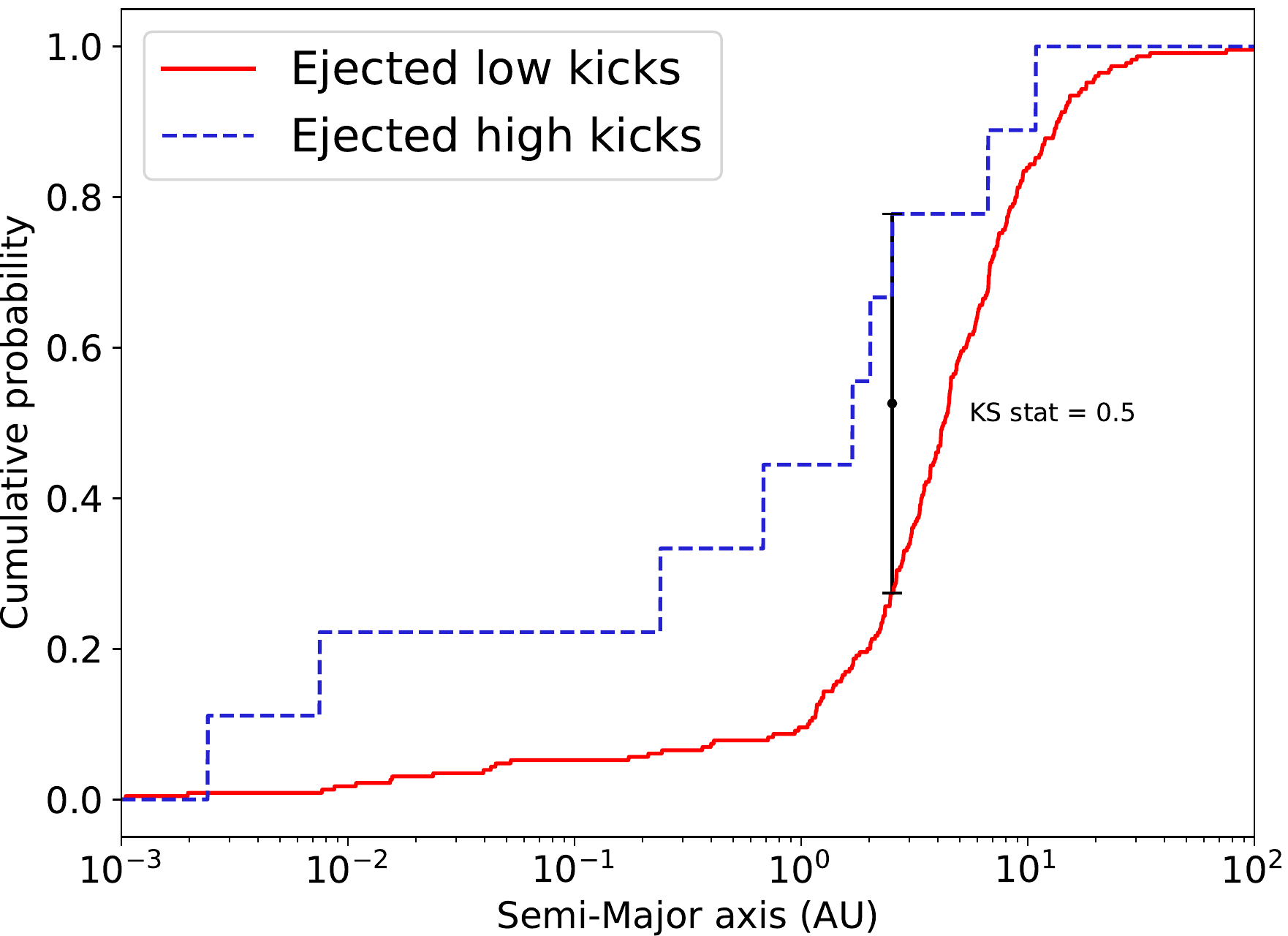}
    \caption{CDF of the semi-major axis at the point of ejection, plotted on a log scale. The red curve shows the distribution for models with the smaller natal kicks, and the blue curve shows the distribution for models with the larger natal kicks. The KS statistic is ~0.5, with a p-value of 0.02.}
    \label{fig:sep_kick_split}
\end{figure}

Figure \ref{fig:sep_kick_split} displays the CDFs of ejected BHB separations for the two kick models. Although the differences between the distributions are statistically significant, this is unlikely to significantly affect results concerning the entire set of ejected systems due to the small number of ejected BHBs from these high kick models (9 out of 239). All of these 9 escaping binaries are ejected well after the formation of all BHs in their host cluster, meaning none are ejected due to a natal kick.

\section{Statistical definitions}
\label{appendix:stats}
\begin{itemize}
    \item \textbf{Savitzky - Golay filter:} The Savitzky - Golay filter is a digital filter used to smooth a set of data points, increasing the precision of the data without altering the signal tendency. For each data point, the filter fits a polynomial to a window of adjacent data points by a method of least squares \citep{1964AnaCh..36.1627S}.  
    \\
    \item \textbf{Outliers:} An outlier is any data point that appears to be outside the general pattern of the data. Although outliers can occur in any legitimate dataset, they are unlikely and so usually indicate some sort of error \citep{moore1993introduction}. There are multiple definitions used to define outliers, with the most common being the so called "Tukey's fence": any data 1.5 times the interquartile range above the third quartile or below the first quartile, \emph{i.e} outside the range $\left[Q_1 - 1.5(Q_3 - Q_1), Q_3 + 1.5(Q_3-Q_1)\right]$ is considered an outlier \citep{tukey}. 
    \\
    \item \textbf{Pearson's chi-square test:} Pearson's chi-square test is a statistical test used on categorical data to determine how likely the observed difference between data sets is purely due to chance. This is often used on a set of events, each corresponding to an outcome of a categorical variable. For example, one could have a set of dice rolls and test if the six-sided die is fair, where the categories are the outcome of the roll. The test statistic is a $\chi^2$ value, which is compared to a $\chi^2$ frequency distribution with the same degrees of freedom as the data. This allows for the null hypothesis (that there is no relation in the frequency of data between categories) to be rejected or supported at a given confidence level \citep{doi:10.1080/14786440009463897}.
    \\
    \item \textbf{Kolmogorov–Smirnov test:} The two-sided Kolmogorov–Smirnov test is a non-parametric statistical test used to compare two data samples to determine if they differ significantly, with the null hypothesis that they are sampled from the same underlying distribution. The test statistic is simply the maximal distance between the two sample cumulative distribution functions. The test accounts for the size of each sample and makes no assumption on the distributions \citep{1987MNRAS.225..155F, stephens1974edf}. 
    \\
    \item \textbf{Spearman rank-order correlation test:} Spearman rank-order correlation test is a non-parametric statistical test used to determine if the correlation between two variables is statistically significant. It assesses both the direction and strength of the monotonic relationship between the two variables and so can be used to test for non-linear correlation \citep{mcdonald2009spearman}.

\end{itemize}

\bibliographystyle{pasa-mnras}
\bibliography{bibliography}

\end{document}